\documentclass[12pt]{iopart}
\usepackage{epsfig}
\usepackage{amsfonts}
\usepackage{psfrag}

\newcommand{\beqn}   {\begin{eqnarray}}
\newcommand{\eeqn}   {\end{eqnarray}}
\newcommand{\ba}   {\begin{eqnarray}}
\newcommand{\ea}   {\end{eqnarray}}

\begin{document}

\title{Intersecting Brane Worlds --
A Path to the Standard Model?}

\author{Dieter L\"ust
\footnote[3]{luest@physik.hu-berlin.de, luest@mppmu.mpg.de}
}

\address{Institut f\"ur Physik, Humboldt-Universit\"at 
zu Berlin, Newtonstr. 15, D-12489 Berlin}

\address{and}

\address{Max-Planck-Institut f\"ur Physik, F\"ohringer Ring 6, D-80805 M\"unchen}

\begin{abstract}
In this review we describe 
the general geometrical framework of brane world constructions 
in orientifolds of type IIA string theory with D6-branes 
wrapping 3-cycles
in a Calabi-Yau 3-fold.
These branes generically intersect 
in points on the internal
space, and the patterns of intersections govern the chiral fermion 
spectra.
We discuss how the open string spectra in intersecting brane
models are constructed, how the Standard Model can be embedded, and also
how supersymmetry can be realized in this class
of string vacua.  After the general considerations
we specialize the discussion to the case of orbifold
backgrounds with intersecting D6-branes and to 
the quintic Calabi-Yau manifold.
Then, we discuss parts of
the effective action of intersecting brane world models. Specifically
we compute from the Born-Infeld action of the wrapped D-branes
the tree-level, D-term scalar potential,
which is important for the stability of the considered backgrounds as well
as for questions related to supersymmetry breaking.
Second, we review the recent computation concerning of
gauge coupling unification and also 
of one-loop
gauge threshold corrections in intersecting brane world models.
Finally we also discuss some aspects of proton decay in intersecting
brane world models.
\end{abstract}



\maketitle

\section{Introduction}

Without any doubt, one important 
goal of superstring theory is to find an
embedding of the Standard Model into a unified
description of gravitational and gauge forces.
However there are several obstacles on the way that one has to solve
in order to achieve this goal:

\begin{itemize}
\item How to derive the precise SM spectrum?

\item How to determine the precise SM couplings?

\item How to break space-time supersymmetry?

\item How to fix the  values of the moduli?

\item How to select the ground state from an (apparent) huge vacuum degeneracy?

\item How to describe the cosmological evolution of the universe?

\end{itemize}

In this review we describe 
the general geometrical framework of brane world constructions 
in orientifolds of type II string theory with D-branes 
wrapping certain homology cycles
in a Calabi-Yau 3-fold.
The branes generically intersect 
each other in the
internal space, and the patterns of intersections govern the chiral fermion 
spectra. 
%and issues of gauge and supersymmetry breaking in the low energy 
%effective gauge theory on their world volume. 
These so-called intersecting brane worlds 
\cite{Blumenhagen:2000wh}--\cite{Blumenhagen:2003su}\footnote{For
%Blumenhagen:2000wh,Angelantonj:2000hi,
%Aldazabal:200dg,Blumenhagen:2000ea,Ibanez:2001nd,Forste:2001gb,
%Blumenhagen:2001te,Cvetic:2001tj,Cvetic:2001nr,Bailin:2001ie,
%Cremades:2002cs,Kokorelis:2002ip,Blumenhagen:2002wn,Uranga:2002pg,
%Blumenhagen:2002gw,Cremades:2003qj,Lust:2003ky,Blumenhagen:2003vr,
%Cvetic:2003ch,Abel:2003fk,
some reviews on many details of open string constructions
see \cite{Angelantonj:2002ct,Ott:2003yv,Goerlich:2004zs}.}
have proven to be a candidate framework of model building which offers 
excellent opportunity to answer some of the
above questions. Namely
in these string compactifications the Standard Model particles correspond
to open string excitations which are located at the various intersections
of the D-branes in the internal 6-dimensional space. 
At the moment, 
type IIA intersecting brane world models with D6-branes, which 
fill the 4-dimensional Minkowski space-time and 
are
wrapped around internal 3-cycles, provide the most promising 
approach to come as close as possible to the Standard Model.\footnote{In a
T-dual mirror description these models correspond to type IIB orientifolds
including D9-branes with magnetic F-flux turned on,
see also the earlier work by \cite{Bachas:1995ik}.}
The fermion spectrum is determined by 
the intersection numbers of the relevant 3-cycles in the internal 
space, as opposed for instance to the older approaches involving heterotic 
strings, where the number of generations was given by the Euler characteristic 
in the simplest case. Indeed it is possible to construct
intersecting brane world models with Standard Model
gauge group and intersection numbers corresponding to three generations
of quarks and leptons.

In a second step, 
going beyond these topological data, 
the computation of the effective
interactions of the light (open) string modes is of vital
importance in order to confront eventually the intersecting brane
world models with experiment. In
particular, the knowledge of the effective
scalar potential 
 \cite{Blumenhagen:2001te,Blumenhagen:2002wn}
is needed to discuss the question of the stability
of intersecting brane configurations and also to know the values
for the various moduli fields.
The moduli dependent gauge coupling constants
\cite{Cremades:2002te,Lust:2003ky}
are essential to get precise informations on the low-energy values
of the Standard Model gauge couplings and the possibility
of gauge coupling unification \cite{Blumenhagen:2003jy}.
The computation of N-point amplitudes is relevant for the
effective Yukawa couplings 
\cite{Cremades:2002cs,Cremades:2003qj,Cvetic:2003ch},
for quartic fermion interactions 
and flavor changing neutral currents 
\cite{Abel:2003fk,Abel:2003vv,Abel:2003yx,Abel:2003yh} as well as for the 
proton decay \cite{Klebanov:2003my}
in intersecting brane world models.
Finally one also needs the effective K\"ahler metric
of the matter fields for the correct normalization of these fields and
also to derive soft supersymmetry breaking parameters
\cite{Kors:2003wf}.

In this work we will review the main aspects of the construction
of intersecting brane world models as well as of the computation
of the tree-level scalar potential and the gauge coupling constants, including
one-loop threshold
corrections.
The further plan of the paper
is like follows:
in the next chapter we will discuss the open string spectrum on
intersecting D-branes, starting with 
local intersecting D-brane constructions in flat space-time,
the question of space-time supersymmetry and
compactifications on Calabi-Yau spaces, i.e.
the embedding of intersecting branes
into a compact CY-manifold. 
Some emphasis is given to the question on the realization of space-time
supersymmetry and the related issue of the stability of
the D-brane configurations, which follows from the minimization
of the associated scalar potential.
As specific examples we will discuss in some detail toroidal and orbifold
models, and also the quintic Calabi-Yau 3-fold.
In the third section we will turn to some
phenomenological issues, namely the 
question of gauge coupling unification.
We will also discuss how to get some quantitative information on 
proton decay in intersecting brane world models.

\section{Intersecting Brane World Models}

\subsection{Intersecting D-branes in flat space-time}
\label{flat}

A lot of the recent progress in type II string physics was made possible
due the discovery of D-branes  \cite{Polchinski:1995mt}.
D(p)-branes are higher(p)-dimensional topological defects, i.e.
hypersurfaces, on which open strings are free to move.
They have led to several new insights:
\begin{itemize}

\item Non-Abelian gauge bosons arise as open strings on the world volumes $\pi$
of the D-branes  \cite{Witten:1995im}. This is the starting point
of the brane world models in type II and type I string models.

\item Chiral fermions arise open strings
living on the intersections of two D-branes \cite{Berkooz:1996km}. 
It follows that the 
number of families is determined by the intersection numbers 
between the brane world volumes $\pi_a$ and $\pi_b$:
\begin{eqnarray}
N_F=I_{ab}\equiv\#(\pi_a\cap\pi_b)\equiv \pi_a\circ\pi_b
\end{eqnarray}

\item D-branes correspond to non-trivial gravitational backgrounds;
this observation led to the famous correspondence between
(boundary) Yang-Mills gauge theories and (bulk) gravitational
theories, the 
AdS/CFT correspondence   \cite{Maldacena:1997re}.

\end{itemize}

First we consider flat D-branes in Minkowski space ${\mathbb R}^{1,9}$.
The
simplest D-brane configuration is certainly just one single Dp-brane:
\vskip0.4cm

\hskip6cm
\epsfig{file=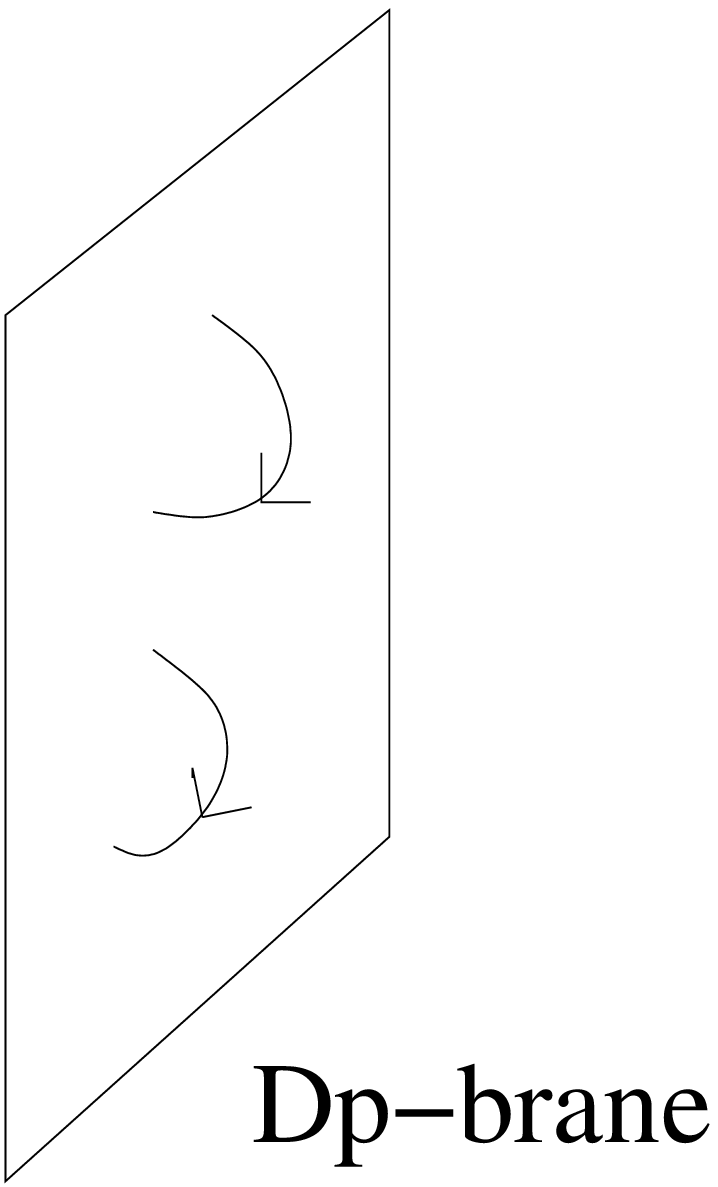,width=2.9cm}

The resulting massless open string spectrum
on the D-brane world volume is then given by one $U(1)$ gauge boson
plus neutral fermions plus scalar fields
which corresponds to a supersymmetric $U(1)$ gauge theory in $p+1$ dimensions
with 16 supercharges.
The effective gauge field action 
for the gauge field strength $F$ on the D-brane
world volume 
\cite{Bachas:1995kx}
contains
the Dirac-Born-Infeld Lagrangian, originating from the tension
of the D-brane, plus a topological Chern-Simons term, which describes
the coupling of the D-brane to the Ramond gauge field potential
$C_{p+1}$ because of its non-vanishing Ramond charge:
\beqn
{\cal S}_{\rm eff} =
\int_{\pi}
     dx^{p+1}\ \big( \underbrace{{\cal L}_{\rm DBI}
(g,{F},\phi)}_{\rm Tension} +
                             \underbrace{{\cal L}_{\rm CS}
({F},C_{p+1})}_{\rm Charge} \big)\, ,
\eeqn
\begin{eqnarray}
S_{DBI}=
\tau_p\int{\rm d}^{p+1}x\sqrt{\det(g_{\mu\nu}+\tau^{-1}F_{\mu\nu})}
=\biggl({M_{\rm string}^{p-3} 
\over g_{\rm string}}\biggr)\int{\rm d}^{p+1}x\, F_{\mu\nu}^2+\dots\, .
\end{eqnarray}
($\tau$ is the string tension,
$\tau_p$ denotes the tension of a Dp-brane, and $\phi$ is the dilaton
field.)

\vskip0.3cm\noindent
Next consider a configuration of a stack of N+M parallel Dp-branes:
\vskip0.4cm

\hskip6cm
\epsfig{file=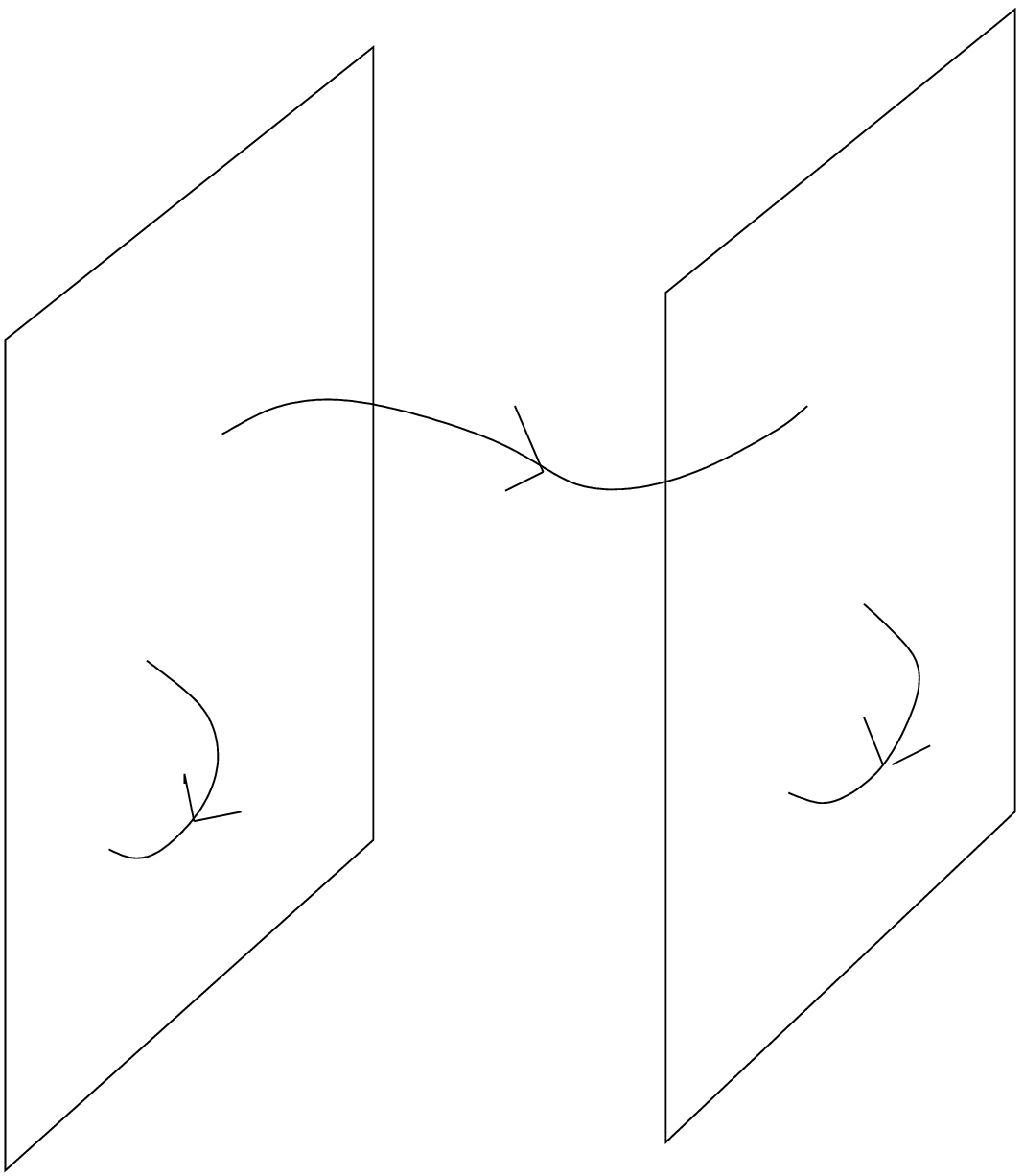,width=2.9cm} 

\noindent
The massless open string spectrum leads to a
${\cal N}=4$ supersymmetric $U(N+M)$ gauge theory in $p+1$ dimensions.
Next we start to simply rotate M D-branes by certain angles 
$\Phi^I_{ab}$ in some directions $I$ with respect to the left-over stack
of N Dp-branes. In other words we are considering
 the case of a stack of N Dp-branes
intersected by another stack of M Dp-branes: 

\vskip0.4cm

\hskip1.5cm
{
\psfrag{a}[bc][bc][.7][0]{$a$} 
\psfrag{b}[bc][bc][.7][0]{$b$} 
 \epsfxsize=4cm
 \epsfysize=4cm
\epsfbox{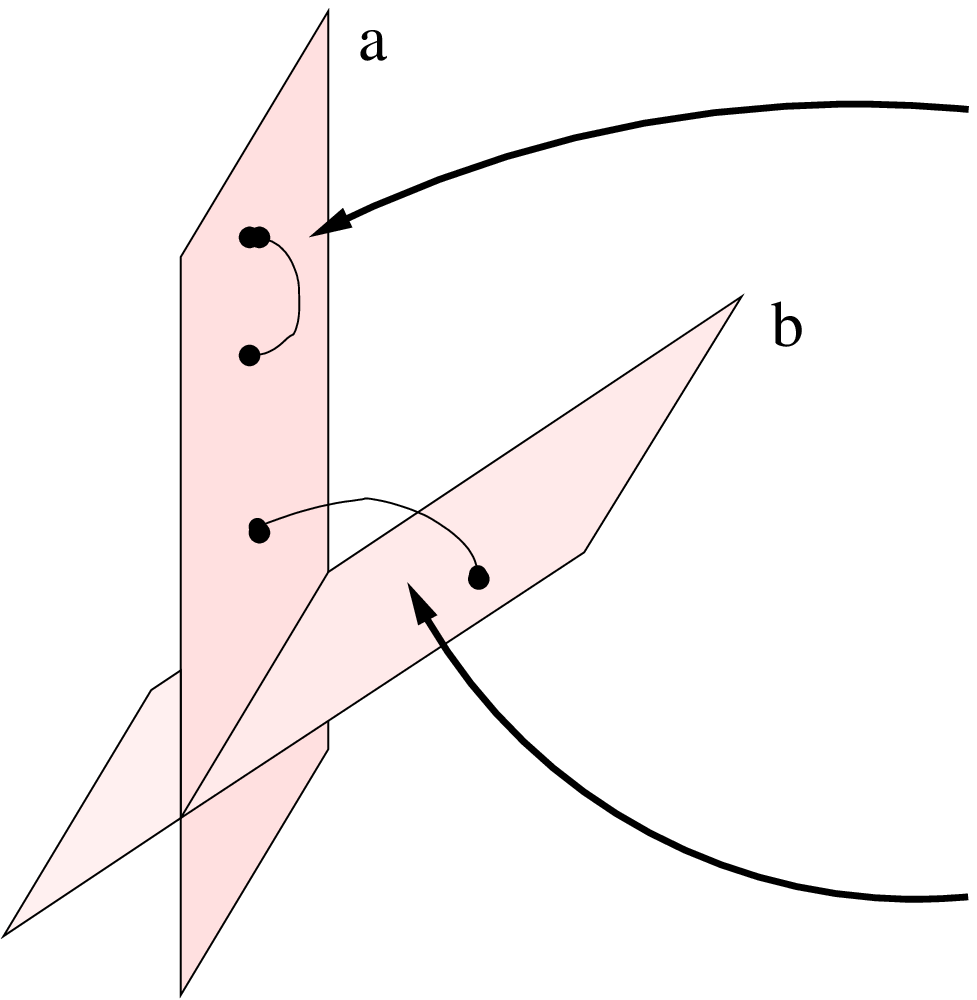}
}
%%%%%%%%%%%%%%%%%%%%%%%%%%%%%%%%%%%%%%%%%%%%%%%%%%%%%
%
%%%%%%%%%%%%%%%%%%%%%%%%%%%%%%%%%%%%%%%%%%%%%%%%%%%%%

\vskip-4cm
\hskip5.7cm
Gauge bosons in {\bf \rm adj.}   

\vskip2.5cm
\hskip5.8cm
Chiral matter in {\bf $(N,\bar M)$}

\vskip1cm\noindent
The open string spectrum on these intersecting branes contains
the following fields  \cite{Berkooz:1996km}:

\vskip0.2cm

\noindent
(i) ${\cal N}=4$ gauge bosons in adjoint representation of $U(N)\times U(M)$.

\noindent
(ii) Massless
fermions in the chiral $(N,\bar M)$ representation.

\noindent
(iii) In general massive scalar fields,
again in the $(N,\bar M)$ representation.

\vskip0.2cm

\noindent
The latter two fields originate from open strings stretching from
one stack of Dp-branes to the other one.
Since the scalar fields are in general massive, such a intersecting
D-brane configurations generically breaks all space-time
supersymmetries.
This supersymmetry breaking
manifests itself as the a
massive/tachyonic scalar ground state with mass:
\beqn
M_{ab}^2 = \frac{1}{2} \sum_{I} \Delta \Phi^I_{ab} - 
{\rm max}\{ \Delta\Phi_{ab}^I \}\, .
\eeqn
($\Phi^I_{ab}$ is the angle between stacks $a$ and $b$ in some
spatial plane $I$.)
Only if the intersection angles take very special values, some
of the scalars become massless, and some part of space-time
supersymmetry gets restored. 
Specifically consider two special flat supersymmetric D6-brane configurations,
as shown in the following figure:

\vskip0.2cm
\begin{center}
\epsfig{file=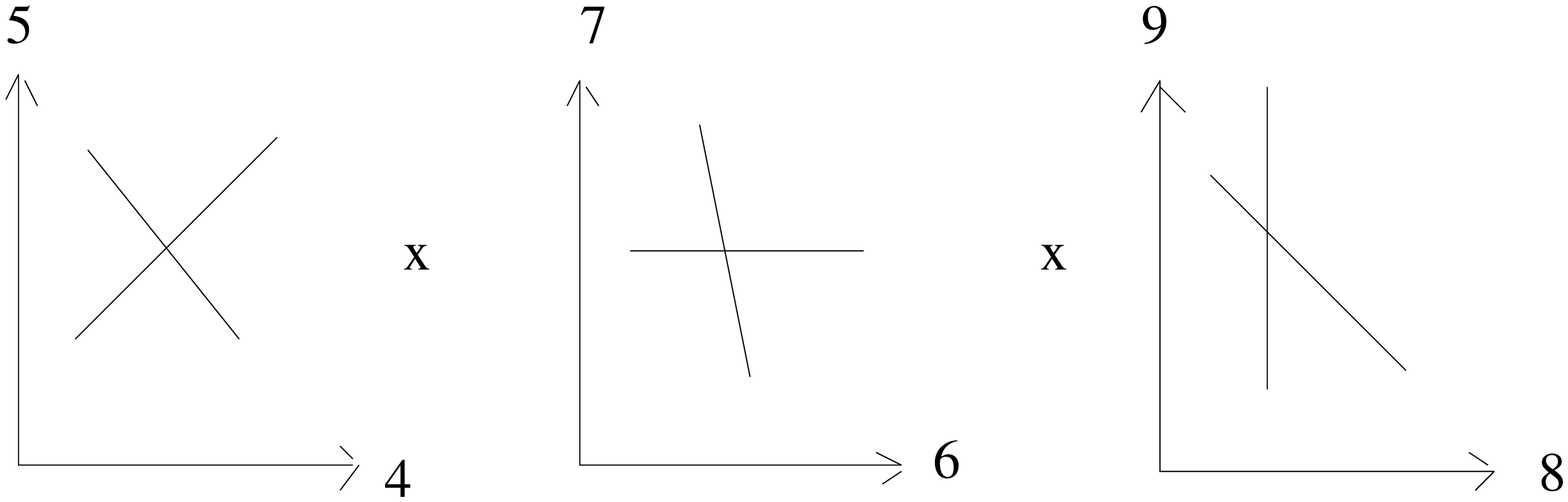,width=13cm} 
\end{center}
\vskip-3.7cm
\hskip2.6cm $\Phi^1$ \\  
%\vskip-0.4cm 
${~}$
\hskip7.2cm $\Phi^2$ \\ \vskip-1cm  ${~}$     \hskip11.2cm $\Phi^3$

\vskip3.5cm

\noindent
Supersymmetry now gets restored for the following choice
of angles:
\begin{itemize}

\item
2 D6-branes, with common world volume in the 123-directions, 
being parallel
in the 4-5, 6-7 and 8-9 planes:
%\beqn
\begin{center}
1/2 BPS (${\cal N}=4$ SUSY): $\Phi^1=\Phi^2=\Phi^3=0$\, .
\end{center}

\item
2 intersecting D6-branes, with common world volume in the 123-directions, 
and which intersect in 4-5 and 6-7 planes, being parallel
in 8-9 plane:
%\beqn
\begin{center}
1/4 BPS (${\cal N}=2$ SUSY): $\Phi^1+\Phi^2=0$, $\Phi^3=0$\, .
\end{center}

\item
2 intersecting D6-branes, common in 123-directions, 
intersecting in 4-5, 6-7 and 8-9 planes:

\begin{center}
1/8  BPS (${\cal N}=1$ SUSY): $\Phi^1+\Phi^2+\Phi^3=0$\, .
\end{center}

\end{itemize}

\noindent
Now, altering the relative intersection angles by some amount,
the open string scalar field will become massive. In case 
$M_{ab}^2>0$ the D-brane configuration is 
a non-supersymmetric, stable non-BPS state 
(possible NS tadpoles,
which cause further instabilities, will be discussed later).
However 
in case the open string scalar is tachyonic ($M_{ab}^2<0)$,
the 2 different branes are unstable and will recombine into a single,
smooth,  supersymmetric D-brane which interpolates between the
intersecting branes,
as can be seen in the next figure:
\begin{center}
\epsfig{file=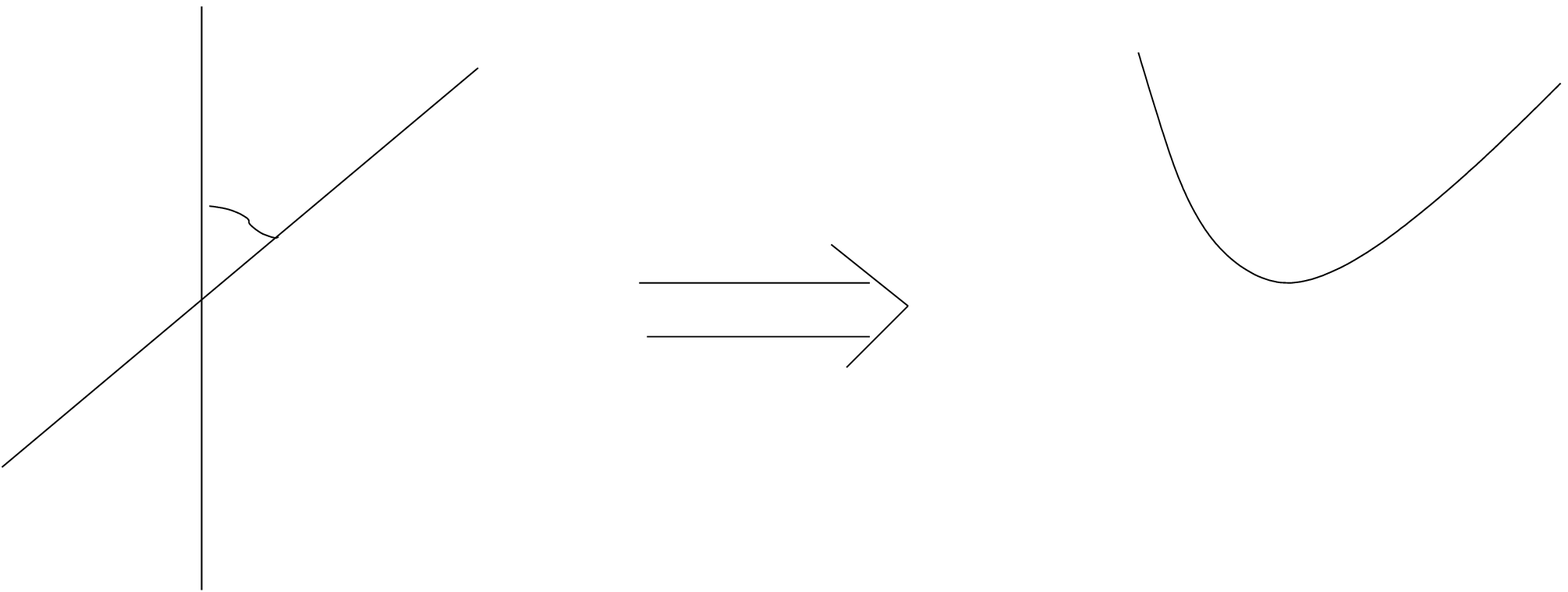,width=13cm} 
\end{center}

\noindent
This process of brane recombination -- the  so-called tachyonic
Higgs effect --  can be described by looking at the effective world volume 
field theory
on the intersecting  D-branes 
\cite{Ohta,Witten:2000mf,Blumenhagen:2000eb,Ohta1,
Erdmenger:2003kn,Epple:2003xt}.
In the world volume field theory
the 
vacuum structure is determined by the D- and F-flatness conditions, which
in the brane picture correspond to the
 geometric calibration conditions.
Intriguingly, the Standard Model Higgs effect might be
realized by tachyon condensation on intersecting D-branes
\cite{Aldazabal:2000dg,Blumenhagen:2001te},
and also in cosmology
 intersecting brane worlds with tachyon condensation 
\cite{Herdeiro:2001zb,Burgess:2001vr,Garcia-Bellido:2001ky,Blumenhagen:2002ua,Dasgupta:2002ew,Gomez-Reino02}
have
been used to model early universe inflation and a ``graceful
exit'' from the inflationary period.

Let us briefly discuss the tachyonic mass spectrum that arise
for two D-branes, which intersect by a single angle $\Phi$.
Starting from the BPS configuration, i.e. two parallel D-branes, and
then rotating them by an angle $\Phi$, 
one gets the following open string spectrum
\begin{equation}
M^2=\left(-\frac{1}{2}-n\right){\frac{|\Phi |}{\pi\alpha'}}+n\, ,
\qquad n=0,1,2,\dots \label{lowlyingstringspectrum}
\end{equation}
for the lowest mass states. If the intersection angle is small,
one single tachyon
field shows up in the open string spectrum. Hence in this case an
effective Yang-Mills
field theory with a finite number of fields  certainly
provides for an appropriate description of the tachyon dynamics.
However, when the intersecting angle is growing, more and more
string modes will become tachyonic. In particular settings where
the intersection angle $\Phi$ is close to $\pi$, which just
corresponds to the small angle intersection of a brane-antibrane
system, contain a large number of tachyonic modes.
For the case $\Phi=\pi$ the
infinitely many tachyons become tachyonic momentum states, and the
corresponding coincident brane-antibrane pair is a highly unstable
non-BPS state, and does not correspond to a perturbative string
ground state. 
Nevertheless, a number of essentially
non-perturbative phenomena have been realized on a field theory
level, notably brane descent relations \cite{Minahan00b,
Hashimoto02, Sen03}, decay of non-BPS branes \cite{Gibbons00,
Sen02b}, brane-antibrane annihilation \cite{Minahan00b, Sen02c}
and local brane recombination \cite{Hashimoto03}. Finally in 
\cite{Epple:2003xt}
some attempts were made
to generalize these effective potentials to
cases where branes and antibranes intersect each other at a small
angle, i.e. the large angle case of intersecting brane.

\vskip0.3cm

Finally at the end
of this section 
we discuss how a local D-brane description of the Standard Model
looks like, describing a D-brane engineering of
the Standard Model by flat D-branes. Later we have to embed this local
D-brane configuration into a compact 6-dimensional space; this will
lead to further consistency conditions. A 
very economic realization of the Standard Model 
is provided by four stacks of D6-branes in the following way
\cite{Blumenhagen:2000ea,Ibanez:2001nd} (see also \cite{Lambert:1999mc}):

\vskip0.5cm
\begin{center}
\begin{tabular}{|c|c|c|c|}\hline
{Stack a:}& $N_a=3$ & $SU(3)_a\times U(1)_a$
& QCD branes
\\ \hline
{Stack b:} & $N_b=2$ & $SU(2)_b\times U(1)_b$&
weak branes
\\ \hline
{Stack c:}& $N_c=1$&  $U(1)_c$&
right brane
\\ \hline
{Stack d:} & $N_d=1$ & $U(1)_d$&
leptonic brane
\\ \hline
\end{tabular}
\end{center}
\vskip1.5cm

\noindent
The intersection pattern of the four stacks of D6-branes can be
depicted in the next figure. The stack $a$ with $N_a=3$ denotes the color 
branes, responsible for the strong QCD forces
with gauge group $SU(3)$, the stack $b$ 
leads to the electro-weak gauge group $SU(2)_L$, and the weak hypercharge
gauge group $U(1)_Y$ is a suitable linear combination of all four
$U(1)$'s.
The left-handed quarks $Q_L$ correspond to massless open string excitations
stretched between the stacks $a$ and $b$, and therefore transform
in the bifundamental $(3,2)$ representation under $SU(3)\times SU(2)_L$,
and so on for the other matter fields. In order to get the correct
hypercharges
for all SM matter fields, the other two stacks $c$ and $d$ are needed.
\begin{center}
\epsfig{file=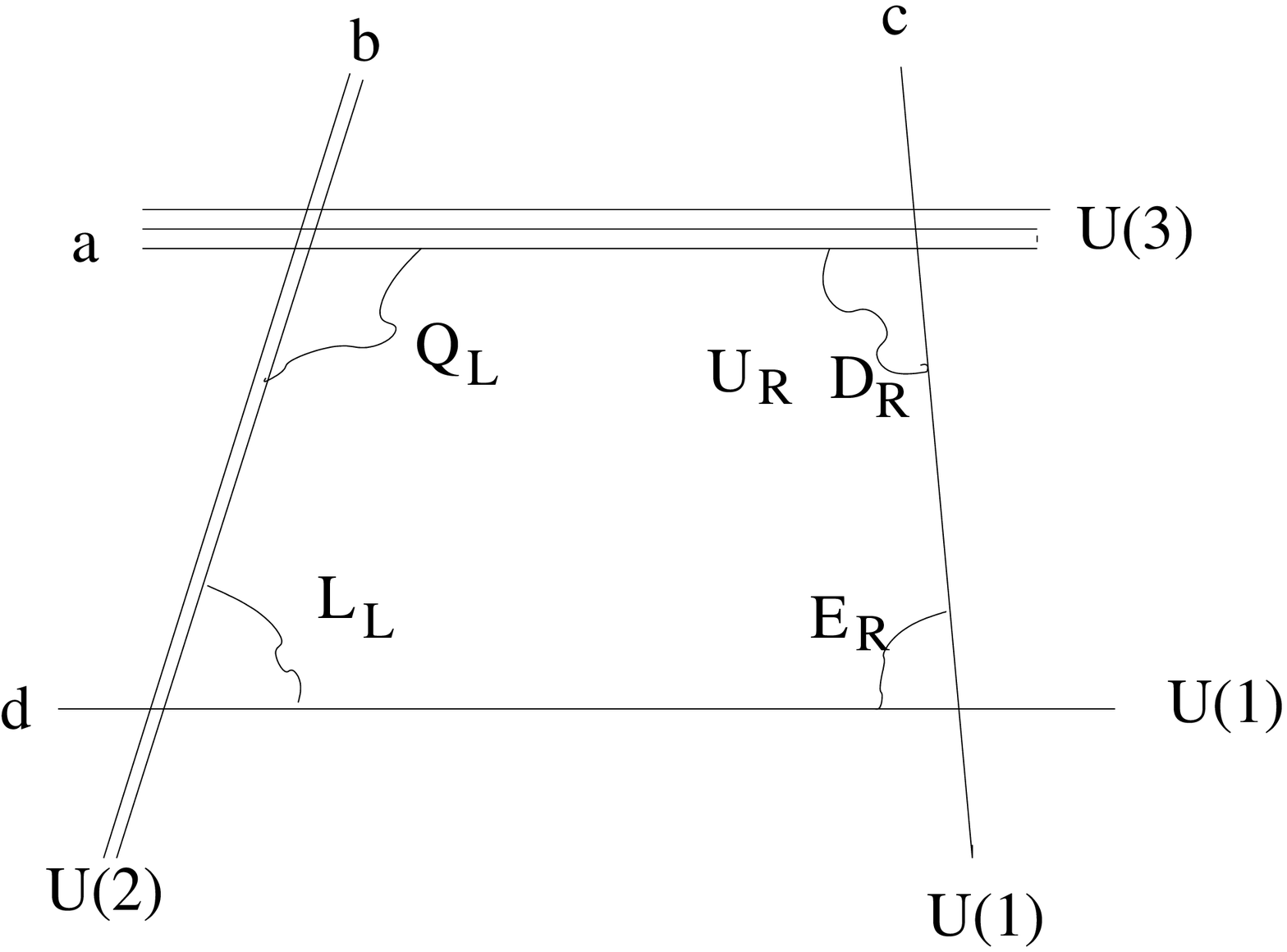,width=11cm} 
\end{center}

\noindent
Space-time supersymmetry is preserved if all four stacks of D6-branes
mutually
preserve the angle conditions among each other.

\subsection{Intersecting D-branes in compact spaces}

 Now we have to embed the intersecting D-branes into a compact space
(we are now following the discussion on intersecting branes 
on Calabi-Yau spaces \cite{Blumenhagen:2002wn,Blumenhagen:2002vp}).
Switching from D-branes in non-compact space to compact spaces
leads to  some important new observations and restrictions which are
essential for model building with intersecting D-branes:

\begin{itemize}
\item Since the D-branes will
be wrapped around compact cycles
of the internal space,
multiple intersections will now be possible.
It follows that the family number will be determined by the topological
intersection numbers of the relevant cycles, around which D-branes are
wrapped.

\item Since
the Ramond charges
of the D-branes cannot `escape'
to infinite, the internal Ramond charges on compact space must cancel
(Gauss law). This is the issue of Ramond tadpole cancellation
which give some strong restrictions on the allowed D-brane
configurations. 

\item Similarly there
is the requirement of cancellation 
of the internal D-brane tensions, i.e
the forces between the D-branes must be balanced.
In terms of string amplitudes, it means that all NS tadpoles must
vanish, namely all NS tadpoles of the closed string moduli
fields and also of the dilaton field. Absence of these tadpoles means
that the potential of those fields is minimized. So this
restriction
imposes a strong
stability problem on non-supersymmetric intersecting D-brane configurations.
On the other hand for supersymmetric D-branes it is automatically satisfied
after Ramond tadpole cancellation.
\end{itemize}

\noindent
In order to meet the two conditions of Ramond and NS tadpole conditions
one in general has to introduce
orientifold planes, which can be regarded as branes of negative
RR-charge and tension.
However we like to emphasize that in intersecting brane models
the tadpoles are in general not canceled locally, i.e.
the D-branes and the orientifold are not lying
on top of each other, but rather only the sum of all charges, integrated
over the entire compact space, has to vanish. This means
that the tadpoles
can be non-locally
canceled by D-branes and orientifolds which are distributed over
the compact space.

Since we are interested in four-dimensional string models in flat
4D Minkowski space-time, we assume that six spatial directions are described
by a compact space  ${\cal M}^{6}$. To be more specific we will consider
a type IIA orientifold background of the form
\begin{eqnarray}
{\cal M}^{10} = ({\mathbb R}^{3,1} \times {\cal M}^{6})/
( \Omega\overline\sigma)\, ,\quad \Omega : {\rm world~ sheet ~parity.}
\end{eqnarray}
%$\Omega$: world sheet parity; 
Here ${\cal M}^{6}$ is a Calabi-Yau 3-fold with a symmetry under 
$\overline\sigma$, the complex conjugation 
\begin{equation}{
\overline\sigma : z_i \mapsto \bar{z}_i,\ i=1,\, ...\, ,3,
}
\end{equation}
in local coordinates $z_i=x^i+iy^i$.
 $\overline\sigma$ is combined with the world sheet parity $\Omega$ 
to form the orientifold projection $\Omega \overline\sigma$. 
This operation is actually a symmetry of the type IIA string 
on ${\cal M}^{6}$. 
Orientifold 6-planes are defined as the 
fixed locus 
\begin{center}
{ ${\mathbb R}^{3,1} \times{\rm Fix}(\overline\sigma)=
{\mathbb R}^{3,1} \times\pi_{O6}
$,} 
\end{center}
where   ${\rm Fix}(\overline\sigma)$ is a supersymmetric (sLag) 3-cycle on
$ {\cal M}^{6}$, denoted by $\pi_{O6}$.
It is special Lagrangian (sLag) and calibrated with 
respect to the real part of the holomorphic 3-form $\Omega_3$.

\vskip0.2cm
\noindent
Next we introduce {D6-branes}  with world-volume
\begin{center}
${\mathbb R}^{3,1} \times\pi_a$, 
\end{center}
i.e. they are
wrapped
around the {supersymmetric (sLag) 3-cycles $\pi_a$ 
and their $\Omega\overline\sigma$ images $\pi_a'$}
of ${\cal M}^6$,
which intersect  in ${\cal M}^6$.
%each other in the
%internal space:
%\newline
%{\B Number of chiral families correspond to the topological intersection
%numbers of the internal 3-cycles!
%}
Then the massless spectrum in general contains the following states:

\begin{itemize}
\item ${\cal N}=1$ supergravity fields in the 10D bulk, since
this orientifold projection truncates the gravitational bulk theory of closed 
strings down to a theory with $16$ supercharges in ten dimensions, 
leading to 4 supercharges and ${\cal N}=1$ in four dimensions, after 
compactifying on the Calabi-Yau.

\item 7-dimensional ${\cal N}=1$ gauge bosons with gauge group 
$G=\prod_{a} U(N_a)$
localized
on stacks of $N_a$
D6-branes wrapped around 
3-cycles $\pi_a$  (codim=3).
In general, the $\pi_a$ are never invariant under $\overline\sigma$
but mapped to image cycles $\pi'_a$. Therefore,
a stack of D6-branes is wrapped on that cycle by symmetry, too.

\item 4-dimensional 
chiral fermions localized on the intersections of the D6-branes
(codim=6).
%The number of chiral families correspond to the topological intersection
%numbers of the internal 3-cycles.
\end{itemize}

\noindent
Since the chiral spectrum has to satisfy some
anomaly constraints, we expect that it is given 
by purely { topological data} (Atiyah-Singer index theorem).
The chiral massless spectrum indeed is completely fixed
by the topological { intersection numbers $I$
of the 3-cycles} of the configuration.
\vskip 0.5cm
\begin{tabular}{lll}

%zeilennormierung
\hspace{2cm} 
&
\hspace{2.5cm} 
&
\\

{\bf Sector} & {\bf Rep.} & {\bf Intersection number I} \\[0.2cm]
\hline
\hline
$a'\,a$
&
$A_a$
&
${1\over 2}\left(\pi'_a\circ\pi_a+\pi_{O6}\circ\pi_a\right)$
\\[0.2cm]

$a'\, a$
&
$S_a$
&
${1\over 2}\left(\pi'_a\circ\pi_a-\pi_{O6}\circ\pi_a\right)$
\\[0.2cm]

$a\,b$
&
$(\overline{N}_a,N_b)$
&
$\pi_{a}\circ \pi_b$
\\[0.2cm]

$a'\, b$
&
$({N}_a,{N}_b)$
&
$\pi'_{a}\circ \pi_b$
\\[0.2cm]
\hline
\end{tabular}
\vskip 0.4cm
\noindent
The non-Abelian gauge anomalies will cancel after satisfying the
tadpole conditions
and 
$U(1)$ anomalies are canceled by a generalized Green-Schwarz
mechanism involving dimensionally reduced RR-forms (see the discussion later
on).

\vskip0.3cm\noindent
Now we are ready to take the following sketchy view on the internal
space:

\begin{center}
   \mbox{ 
         % \begin{turn}{-90}  % Hier k"onnen Sie das Bild drehen 
                             % z.B. um -90 Grad
             \epsfig{file=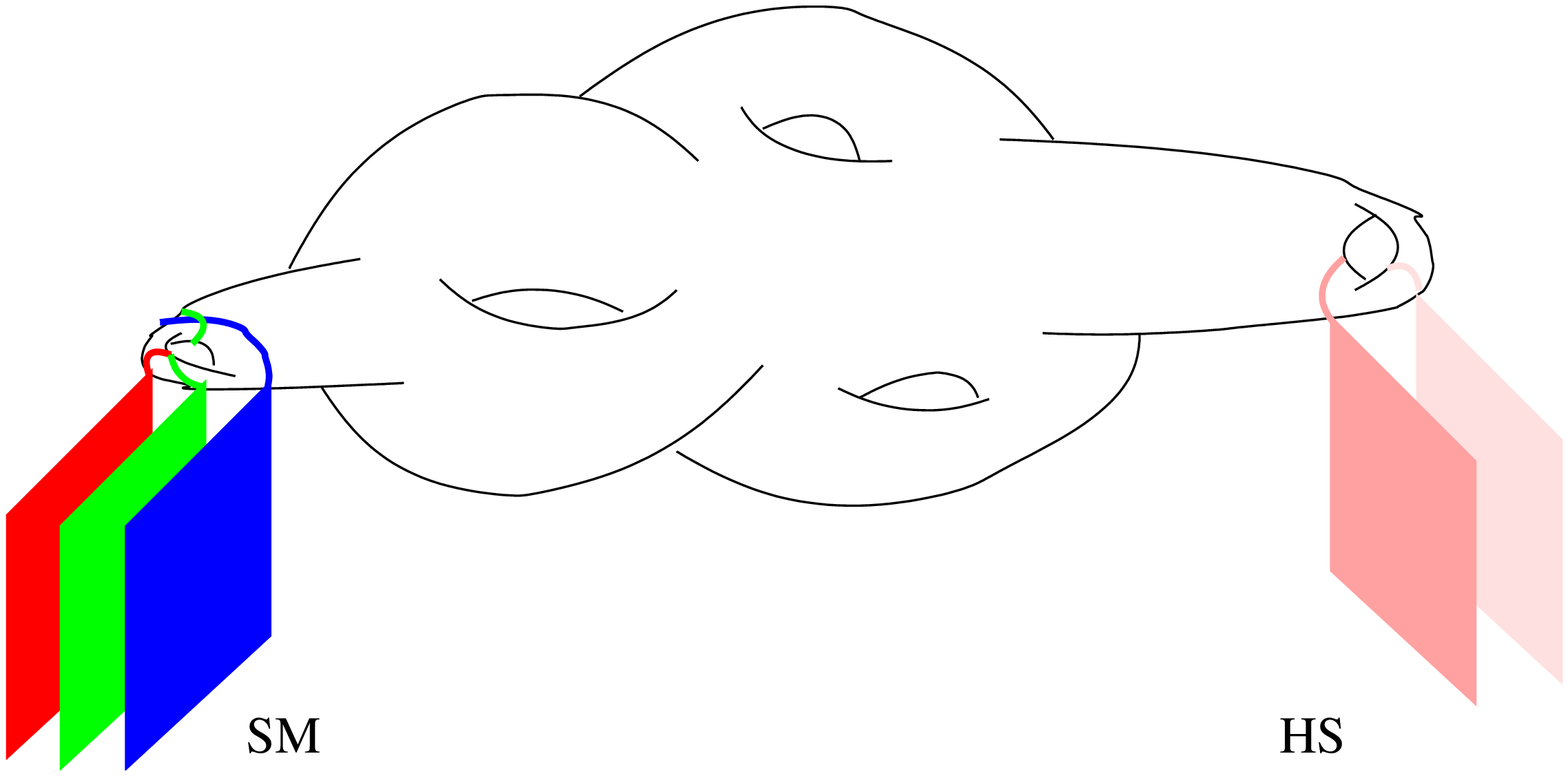,width=15.5cm}
         % \end{turn}
}
   \end{center}

\vskip0.3cm\noindent
Here one sector of intersecting D6-branes engineers the embedding
of the Standard Model, the other sector of D6-branes  corresponds to
a hidden gauge sector which may be required by the tadpole cancellation
conditions. Note that the hidden sector D-branes do not have any intersection
with the Standard Model branes. Therefore there are no matter fields
charged under the Standard
Model and hidden gauge group. It follows that the hidden sector and
the Standard Model sector can only interact with each other by
gravitational effects or by the exchange of heavy string modes.

This kind of scenario allows for at least three possibilities 
for the realization and the breaking of space-time supersymmetry:

\begin{itemize}
\item
The SM-branes are non-supersymmetric. Then the scale of supersymmetry
breaking is of the order of the string scale, which, in
order to solve the hierarchy problem, should be of the order
of the TeV scale:
\beqn
M_{\rm susy}\simeq M_{\rm string}\sim {\cal O}(1~{\rm TeV})
\eeqn
In order to explain
the weakness of
the gravitational force
this scenario needs for large transversal dimensions $R_\perp$ on the 
the internal space  \cite{Arkani-Hamed:1998rs,Antoniadis:1998ig}.

\item 
The SM-branes are mutually supersymmetric
({``local" supersymmetry}), but are non-super\-sym\-metric with respect
to hidden sector branes. In this case one generically deals the
gravity mediated supersymmetry breaking:
\beqn
M_{\rm susy}\simeq{M^2_{\rm string}\over M_{\rm Planck}}\simeq{\cal O}(
1{\rm TeV})
~\Rightarrow ~M_{\rm string}\simeq{\cal O}(10^{11}{\rm GeV})
\eeqn
{Here the
transversal dimensions on the internal space are only moderately enlarged, 
($R_\perp\simeq
{\cal O}(10^9){\rm GeV})$ (see also \cite{Burgess:1998px}).
}
\item
All branes are mutually supersymmetric ({``global" supersymmetry})
Then supersymmetry must be broken by
a dynamical mechanism (e.g. gaugino condensation) in hidden sector:
\beqn
M_{\rm susy}\simeq{M^3_{\rm hidden}\over M_{\rm Planck}^2}\simeq{\cal O}(
1{\rm TeV})
~\Rightarrow ~M_{\rm hidden}\simeq{\cal O}(10^{13}{\rm GeV})
\eeqn

\end{itemize}

\vskip0.3cm
Lets us now come to the consistency conditions for D-branes on compact
spaces.
We start with 
the requirement of 
 RR-charge cancellation.
 For that purpose we need the topological Chern-Simons actions for Dp-branes
and also for the orientifold Op-plane which have
the form    
\cite{Douglas:1995bn,Green:1996dd,
Morales:1998ux,Scrucca:1999uz,Stefanski:1998yx}:
\begin{eqnarray}
\label{cs}
{\cal S}^{({\rm D}p)}_{\rm CS} & = &
\mu_p \int_{{\rm D}p}
{\rm ch}({\cal F})\wedge  
\sqrt{{\hat {\cal A}({\cal R}_T) \over \hat {\cal A}({\cal R}_N)
   }} \wedge \sum_q{C_q} , \nonumber\\ 
{\cal S}^{({\rm O}p)}_{\rm CS} & = &
-2^{p-4} \mu_p \int_{{\rm O}p}
    \sqrt{{\hat {\cal L}({\cal R}_T/4) \over \hat{\cal L}({\cal R}_N/4)
   }} \wedge \sum_q{C_q} .
\end{eqnarray}
(ch$({\cal F})$ denotes the Chern character, 
$\hat{\cal A}({\cal R})$ the Dirac genus of the tangent or normal bundle, 
and the $\hat{\cal L}({\cal R})$ the Hirzebruch polynomial.) 
The physical gauge fields and curvatures are related to the
skew-hermitian ones in (\ref{cs})  by rescaling with $-4i\pi^2\alpha'$. 
These expressions simplify drastically for sLag 3-cycles, where 
${\rm ch}({\cal F})\vert_{{\rm D}p}={\rm rk}({\cal F})$, the other 
characteristic classes become trivial and finally the only contribution in the
CS-term (\ref{cs})   for D6-branes
comes from $C_{7}$. 
Then this action leads
to the following equations of motion for the gauge field $C_7$:
\beqn
{ {1\over \kappa^2}\,
d\star d C_{7}=\mu_6\sum_a N_a\, \delta(\pi_a)+
                    \mu_6\sum_a N_a\, \delta(\pi'_a)
        + \mu_6 Q_6\,  \delta(\pi_{{\rm O}6}),
}
\eeqn 
where $\delta(\pi_a)$ denotes the Poincar\'e dual form of $\pi_a$, 
$\mu_p = 2\pi (4\pi^2\alpha')^{-(p+1)/2}$, and 
$2 \kappa^2= \mu_7^{-1}$.
Upon
integrating over ${\cal M}^6$ one obtains the RR-tadpole cancellation 
as an equation
in homology \cite{Blumenhagen:2002wn}:
\beqn
\label{tadpoles}
\sum_a  N_a\, (\pi_a + \pi'_a)-4\pi_{{\rm O}6}=0.
\eeqn
In principle it involves as many linear relations as there are
independent generators in $H_3({\cal M}^{6},R)$. But, of course,
the action of $\overline\sigma$ on ${\cal M}^{6}$ also induces an action
$[\overline\sigma]$ on the homology and cohomology. In particular,
$[\overline\sigma]$ swaps $H^{2,1}$ and $H^{1,2}$, and 
the number of conditions is halved.

Note that one can show that
the cancellation of
the RR tadpoles implies absence of 
the non-Abelian anomalies in the effective 4D field theory.
However there can be still anomalous $U(1)$ gauge symmetries
in the effective 4D field theory.
These anomalies will be canceled by a Green-Schwarz mechanism
involving RR (pseudo)scalar field. As a result of these interactions
the corresponding $U(1)$ gauge boson will become massive.
Specifically we have to consider 
two relevant couplings in the effective action:
\begin{eqnarray}
\int_{R^{3,1}\times \pi_a}C_5\wedge {\rm Tr}F_a &\sim &
\int_{R^{3,1}}B\wedge F_a\, ,\nonumber \\
\int_{R^{3,1}\times \pi_b}C_3\wedge {\rm Tr}(F_b\wedge F_b) &\sim &
\int_{R^{3,1}}\phi\wedge (F_b\wedge F_b)\, . 
\end{eqnarray}
They lead to the following  diagram which cancels the anomalous
triangle diagram:
\begin{center}
\epsfig{file=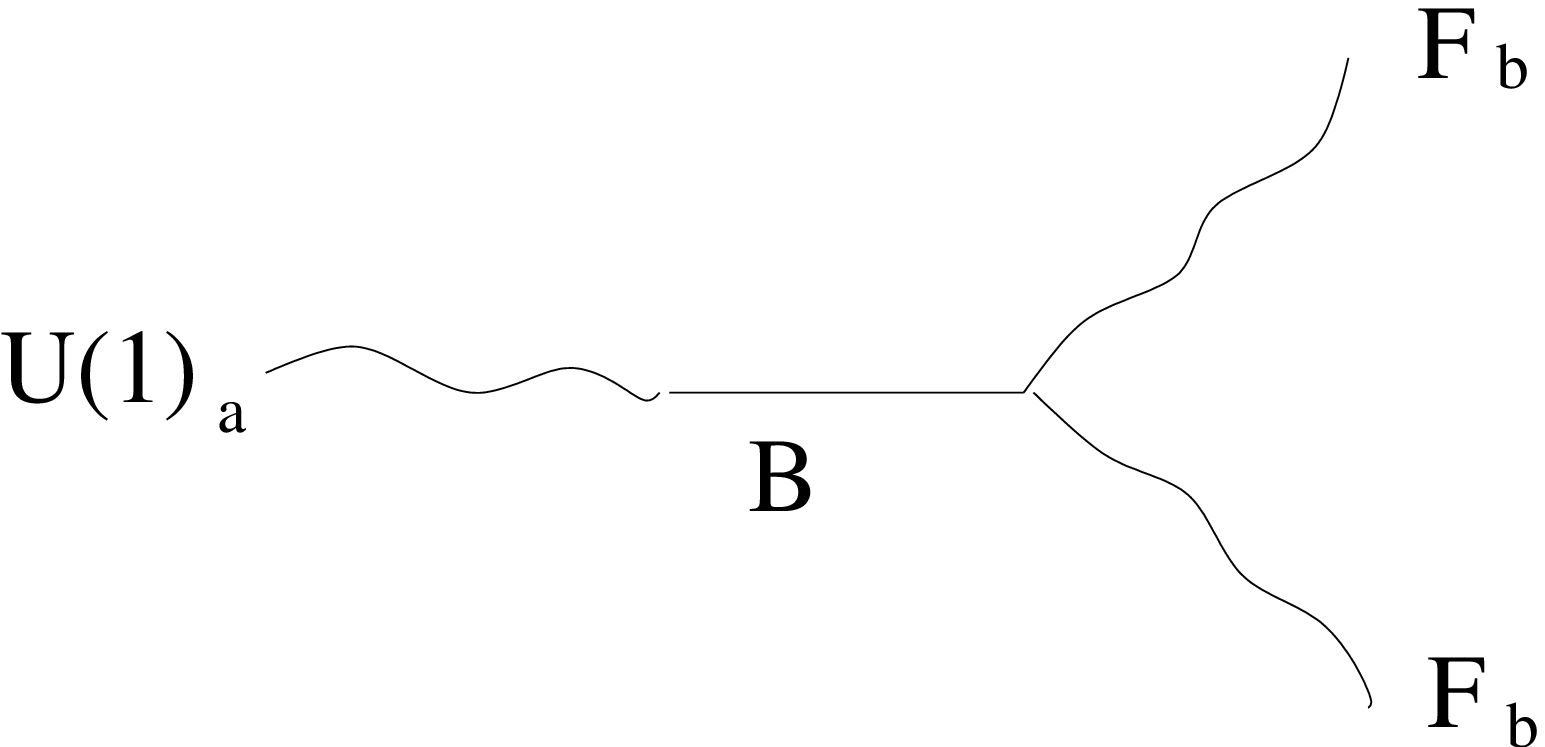,width=7cm} 
\end{center}
Now the condition for an anomaly free $U(1)_a$ is:
\begin{eqnarray}
N_a(\pi_a-\pi_a')\circ \pi_b=0\, .
\end{eqnarray}
Note that 
even an anomaly free $U(1)$ can become massive, if only the $U(1)_a-B$
coupling is present.
The massive $U(1)$ always remains as a global symmetry.

\vskip0.3cm
As the second consistency requirement we now turn to the
stability of the scalar potential, related to
the absence of NS tadpoles.
The tension of the D6-branes and O6-planes introduces a 
vacuum energy 
which is described in terms of D-terms in the language of ${\cal N}=1$ 
supersymmetric field theory. In type IIA
these depend only on the complex structure 
moduli and do not affect the K\"ahler parameter of the background. 
The most general form for such a potential is given by
\begin{equation}
\label{fi}
{\cal V}_{\rm D-term} =
\sum_a {1\over 2g_a^2}\biggl(\sum_i q^i_a|\phi_i|^2+\xi_a\biggr)^2\, ,
\end{equation}
with $g_a$ the gauge coupling of a $U(1)_a$, 
$\xi_a$ the FI parameter, and the
scalar fields $\phi_i$ are the superpartners of some bifundamental 
fermions at the 
intersections. They become massive or tachyonic 
for non-vanishing $\xi_a$, depending on their charges $q^i_a$. 
Due to the positive definiteness of the D-term, 
${\cal N}=1$ supersymmetry will only be unbroken in the vacuum,
if the potential vanishes.

The disc level tension can be determined by 
integrating the Dirac-Born-Infeld effective
action. It is proportional to the
volume of the D-branes and the O-plane, so that the
disc level scalar potential reads
\begin{equation}
\label{sp}
{\cal V}=T_6\, {e^{-\phi_{4}}
\over \sqrt{{\rm Vol({\cal M}^{6})}}}
               \left( \sum_a  N_a \left( {\rm Vol}({\rm D}6_a) + 
      {\rm Vol}({\rm D}6'_a) \right) -4 {\rm Vol}({\rm O}6)\right). 
\end{equation}
The potential is easily seen to be positive semidefinite and its 
minimization imposes conditions on some of the moduli, freezing them 
to fixed values. Whenever the potential is non-vanishing, supersymmetry 
is broken and a classical vacuum energy generated by the net brane tension. 
It is easily demonstrated that the vanishing of ${\cal V}$ requires all 
the cycles wrapped by the D6-branes to be calibrated with respect to the same 
3-form as are the O6-planes. In a first step, just to conserve supersymmetry 
on their individual world volume theory, the cycles have to be 
calibrated at all, which leads to  \cite{Blumenhagen:2002wn} 
\begin{equation}{
{\cal V}=T_6\, e^{-\phi_{4}} \left(
\sum_a{N_a \left| \int_{\pi_a} \widehat\Omega_3 \right|} +
 \sum_a{N_a \left| \int_{\pi'_a} \widehat\Omega_3 \right|} 
-4 \left| \int_{\pi_{{\rm O}6}} \widehat\Omega_3 \right |\right) .
}
\end{equation}

Since $\widehat\Omega_3$ is closed, the integrals only depend on
the homology class of the world volumes of the branes and planes and 
thus the tensions also become topological. If we further demand that 
any single D$6_a$-brane conserves the same
supersymmetries as the orientifold plane the cycles 
must all be calibrated with respect to $\Re(\widehat\Omega_3)$. We can 
then write 
\begin{equation}{
{\cal V}=T_6\,  e^{-\phi_{4}} 
\int_{\sum_a{N_a (\pi_a+\pi_a')
-4\pi_{{\rm O}6}}} \Re(\widehat\Omega_3) .
}
\end{equation}
In this case, the RR charge and NSNS tension cancellation is 
equivalent, as expected in the supersymmetric situation.

To apply eq.(\ref{fi}) we have to use the properly normalized gauge coupling 
\begin{equation}
\label{gc}
{1\over g_{U(1)_a}^2}={N_a\over g_a^2}={N_a\, M_s^3\over (2\pi)^4}\,
                     e^{-\phi_{4}}
                \left| \int_{\pi_a} \widehat\Omega_3 \right|.
\end{equation}
Hence, the FI-parameter $\xi_a$ can be identified as
\begin{equation}{
\xi^2_a={M_s^4 \over 2\pi^2} { \left| \int_{\pi_a} \widehat\Omega_3 \right|-
               \int_{\pi_a} \Re(\widehat\Omega_3) \over
                \left| \int_{\pi_a} \widehat\Omega_3 \right|} ,
}
\end{equation}
which vanishes precisely if the overall 
tension of the branes and planes cancels out, 
i.e. if all are calibrated with respect to the same 3-form. 
Since the FI-term is not a holomorphic quantity one expects
higher loop corrections to the classical potential eq.(\ref{sp}).

Summarizing this discussion on the scalar potential,
we can consider three different scenarios
with respect of the realization of space-time supersymmetry (see above):

\begin{itemize}

\item
{\sl ``Global" ${\cal N}=1$ supersymmetry:} \newline
Here the minima  of ${\cal V}$
with respect
to the complex structure moduli are such that 
all angles are supersymmetric. Therefore all 
angles, being determined by
the complex structure moduli, of the D6-branes are supersymmetric
in the minimum and
conserve the same supersymmetries as orientifold plane, i.e. all D6-branes
be calibrated with respect to $\Re(\widehat\Omega_3)$. Furthermore
the vacuum energy vanishes in the supersymmetric
minimum ${\cal V}_{\rm min}=0$.

\item
{\sl ``Local" ${\cal N}=1$ supersymmetry:} \newline
Now the minima of ${\cal V}$ are
 such that only the SM D-branes have mutually
supersymmetric
angles; only the SM D6-branes
conserve the same supersymmetries as orientifold plane, i.e. only SM D6-branes
be calibrated with respect to $\Re(\widehat\Omega_3)$.
However
the hidden sector is in general necessary for RR tadpole cancellation.
Note that stable minima of this type are already non-trivial to find,
since we eventually have to stabilize 
the SM branes with respect to the hidden branes.

\item
{\sl No supersymmetry:} \newline
Finally in this case the minima are such that the
angles of the SM branes are non-supersymmetric. Therefore
in the minimum the SM D6-branes
do not
conserve the same supersymmetries as the orientifold plane. 
%i.e. SM D6-branes
%are not calibrated with respect to $\Re(\widehat\Omega_3)$.
Here stability is in general very difficult to achieve, as we also must
be careful to avoid the process of brane recombination to a supersymmetric
scenario. In order to determine the full non-supersymmetric
ground date, non-perturbative (string field theory) methods will be
necessary.

\end{itemize}

In addition to the D-term scalar potential, also a F-term potential
corresponding to an effective superpotential
might be generated. Here
there are strong restrictions known for the contributions that can 
give rise to corrections to the effective ${\cal N}=1$ superpotential 
of a type II compactification on a Calabi-Yau 3-fold with D6-branes and
O6-planes on supersymmetric 3-cycles.
The standard arguments about the non-renormalization of the superpotential by 
string loops and world sheet $\alpha'$ corrections apply. 
The only effects then 
left are non-perturbative world sheet corrections, open and closed 
world-sheet instantons  
\cite{Aganagic:2001nx,Lerche:2002ck}. In general, 
these are related to non-trivial $CP^1$ and 
$RP^2$ with boundary on the O6-plane in the Calabi-Yau manifold 
for the closed strings 
and discs with boundary on the D6-branes for open strings.
In fact, only the latter contribute to the superpotential. 
The typical form for the superpotential thus generated is known, but 
explicit calculations are only available for non-compact models. Usually, they 
make use of open string mirror symmetry arguments. 
In many cases, there is an indication that the non-perturbative 
contributions to the superpotentials 
tend to destabilize the vacuum, and it would be a tempting 
task to determine a class of stable ${\cal N}=1$ supersymmetric
intersecting brane models.

\vskip0.3cm\noindent
Let us now continue with some further aspects of string model
building, namely how to derive the spectrum of the Standard Model.
In fact, there are two 
simple ways to embed the Standard Model 
\cite{Blumenhagen:2000ea,Ibanez:2001nd,Cremades:2003qj}.
Both of them use { four stacks of $D6$-branes}, but
they differ in their realization of the weak $SU(2)$ gauge group
(note that $SP(2)$ and $SU(2)$ are isomorphic groups):
\begin{eqnarray}
A&:&\  U(3)_a\times SP(2)_b \times U(1)_c\times U(1)_d \nonumber \\
B&:&\  U(3)_a\times U(2)_b \times U(1)_c\times U(1)_d.   
\end{eqnarray}
The chiral spectrum of the intersecting brane world
model should be identical to the chiral spectrum 
of the standard model particles.
This {fixes uniquely the intersection numbers} of the
four  
3-cycles, $(\pi_a,\pi_b,\pi_c,\pi_d)$; for all class A models one derives
the following specific intersection assignment (a similar
pattern holds for all class B models):

\vskip 0.5cm
\begin{tabular}{llll}

%zeilennormierung
\hspace{2cm} 
&
\hspace{2.5cm} 
&
\\

{\bf field} & {\bf sector} & {\bf I} & $SU(3)\times SU(2)\times U(1)^3$\\[0.2cm]
\hline
\hline
$Q_L$ & (ab) & 3 & $(3,2;1,0,0)$ 
\\[0.2cm]
$U_R$  & (ac)  & 3 & $(\overline{3},1;-1,1,0)$ 
\\[0.2cm]
$D_R$  & (ac') & 3 & $(\overline{3},1;-1,-1,0)$ 
\\[0.2cm]
\hline
$E_L$  & (db) & 3 & $(1,2;0,0,1)$ 
\\[0.2cm]
$E_R$  & (dc') & 3 & $(1,1;0,-1,-1)$ 
\\[0.2cm]
$\nu_R$  & (dc) & 3 & $(1,1;0,1,-1)$ 
\\[0.2cm]
\hline
\end{tabular}
\vskip 0.4cm\noindent
The {hypercharge $Q_Y$} is given as the following linear combination
of the three $U(1)$'s
\begin{eqnarray}
Q_Y={1\over 3}Q_a-Q_c-Q_d .
\end{eqnarray}

\noindent
Then an intersecting brane world 
model is constructed by the following six steps:  

\noindent
(i) chose a compact toroidal, orbifold or Calabi-Yau manifold  ${\cal M}_6$,

\noindent
(ii) determine the orientifold 6-plane $\pi_{O6}$,

\noindent
(iii) chose four 3-cycles $\pi_{U(3)_a}$, $\pi_{U(2)_b}$, 
$\pi_{U(1)_c}$, $\pi_{U(1)_d}$ for the
four stacks of D6-branes, as well as their orientifold
mirrors,

\noindent
(iv) compute their intersection numbers,

\noindent
(v) ensure that the RR tadpole conditions vanish (possibly by adding
hidden D6-branes),

\noindent
(vi) and finally ensure 
that the  linear combination $U(1)_Y$
remains massless.

\begin{eqnarray}
  \sum_i  c_i\, N_i\, \left(\pi_i-\pi_i'\right)=0 \, ,
\end{eqnarray}
where the $c_i$ define the precise linear combination for $U(1)_Y$.

\vskip0.2cm\noindent
In this way
many non-supersymmetric
intersecting brane world models on tori
\cite{Blumenhagen:2000ea,Ibanez:2001nd}, 
orbifolds \cite{Blumenhagen:2001te},
or the quintic Calabi-Yau manifold 
\cite{Blumenhagen:2002wn,Blumenhagen:2002vp}
with gauge group
 \beqn
G= SU(3)_c \times SU(2)_L \times U(1)_Y \nonumber
\eeqn

\noindent
and 3 families of quark and leptons can be explicitly constructed.
(Some authors also use more than four stack of D6-branes
or different types of D-branes.
For intersecting branes on supersymmetric orientifolds of Gepner models
see \cite{Blumenhagen:2003su,brunnero,blumenhageno}.)
In order to construct models with the spectrum of the ${\cal N}=1$
supersymmetric model (MSSM) also various attempts were made
\cite{Cvetic:2001tj,Cvetic:2001nr,
Blumenhagen:2002gw,Honecker:2003vq,Honecker:2003vw}. 
However
as far as we know no explicit model with the precise
spectrum of the MSSM was yet found. 

Let us first describe in more detail how one constructs toroidal and
orbifold intersecting brane world models. Here
we consider configurations of type II $D6$ branes wrapped on non--trivial 
three--cycles of a six--dimensional torus $T^6$. 
The torus is taken to be a direct product $T_6=\prod\limits_{j=1}^3 T_2^j$
of three two--dimensional tori $T_2^j$ with radii $R_1^j,R_2^j$ and angles 
$\alpha^j$ w.r.t.  the compact dimensions 
with coordinates $x^j$ and $y^j$. 
The K\"ahler and complex structure moduli of
these tori are defined as usual: 
%{\bf This is not standard definition for $T$ in type II !.
%Shall we be more precise. St. will take care of it.}
\begin{equation}
\label{moduli}
U^j=\frac{R_2^j}{R_1^j}e^{i\alpha^j}\ \ ,\ \
T^j=b^j+iR_1^jR_2^j\sin\alpha^j\, ,
\end{equation}
with the torus $B$--field $b^j$.
Furthermore, the three--cycle is assumed to 
be  factorizable into a direct product of 
three one--cycles, each of them wound around a torus $T_2^j$ with the wrapping
numbers $(n^j,m^j)$ w.r.t. the fundamental 1--cycles of the torus.
Hence the angle of the $D6$--brane with the $x^j$--axis is given by
\begin{equation}
\label{angles}
\tan \Phi^j=\frac{m^jR^j_2}{n^jR^j_1}\, .
\end{equation}
Generally, two branes with wrapping numbers $(n^j_a,m^j_a)$ and
$(n^j_b,m^j_b)$, are parallel in the subspace $T_2^j$, 
if their intersection number
\begin{equation}{
I_{ab}^{j}= n^j_am^j_b-n^j_bm^j_a  }
\end{equation}
w.r.t.  this subspace vanishes, $I_{ab}^{j}=0$. 

As explained before stability of a given D-brane configuration requires
that all NS tadpoles vanish. 
The relevant massless fields in the NS sector are the four-dimensional
dilaton and the complex structure moduli $U^i$.
One method is to extract the NS tadpoles from the
infrared divergences in the tree channel Klein-bottle, annulus and
M\"obius-strip amplitudes, the open string one-loop diagrams. 
Adding up these three
contributions one can read off the disc tadpoles. Another method
is to compute
the
corresponding scalar potential from the
Born-Infeld action of
the wrapped D6-branes; in the string frame it takes the following
form $(U^i=U^i_1+iU^i_2$)  \cite{Blumenhagen:2001te}:
\begin{eqnarray}
{\cal V}&=& e^{-\phi}\, 
%\left
 \sum_{a=1}^K  N_a\, \prod_{i=1}^3 
   \sqrt{\left( n_a^i\, \sqrt{U_2^i}\right)^2 + 
                          \left( (m_a^i+U_1^i\, n_a^i)\, 
                {1\over \sqrt{U_2^i}}\right)^2} \nonumber \\ 
          &-& 16
 e^{-\phi}
\, \prod_{i=1}^3 \sqrt{U_2^i} 
%\right
\end{eqnarray}
The NS tadpoles are simply the first derivatives
of ${\cal V}$ with respect to the scalar fields:
$\langle \phi \rangle \sim {\partial {\cal V}
 \over \partial \phi}$, 
$\langle U^i \rangle \sim {\partial V \over \partial U^i}$.
So we clearly see that the NS tadpoles vanish in case the potential
is extremized. For supersymmetric  
minima the potential in addition vanishes, and all D-branes
have supersymmetric angles at the minimum.
Then, at the minimum of ${\cal V}$, 
all the  3-cycles on the torus including
the orientifold 3-cycle, constructed in the way described above, 
are supersymmetric, i.e. sLag's, for all values of $n_a^i$ and $m_a^i$.
For supersymmetric orbifold models one therefore has to check whether they
conserve the same supersymmetries as the orientifold plane.
Specifically one obtains the following conditions on the angles $v_a^j$
of a stack of supersymmetric $D6_a$ branes:

\vskip0.2cm
\noindent (i) ${\cal N}=4$ sectors: $v_a^1=v_a^2=v_a^3=0$.

\vskip0.2cm
\noindent (ii) ${\cal N}=2$ sectors: $v_a^i=0$ w.r.t.
the i-th plane and $v_a^j\pm v_a^l=0$.

\vskip0.2cm
\noindent (iii) ${\cal N}=1$ sectors: $v_a^1\pm v_a^2\pm v_a^3=0$.

\vskip0.2cm

In non-supersymmetric models stability and tadpole cancellation is
much harder to acquire.
The potential displays the usual runaway behavior
one often encounters in non-supersymmetric string models. 
The complex structure is dynamically pushed to the degenerate limit,
where all branes lie along the $x_i$ axes and the $y_i$ directions
shrink, keeping the volume fixed. Put differently, the positive tension of the
branes pulls the tori towards the $x_i$-axes. 
Apparently, this has dramatic consequences for all toroidal intersecting
brane world models. 
They usually require a tuning of parameters at tree-level and  
assume the global stability of the background geometry as given by the closed
string moduli. 
If at closed string tree-level one has arranged the radii
of the torus such that open strings stretched between D-branes
at angles are free of tachyons, dynamically the system flows towards
larger complex structure and will eventually reach a point where
certain scalar fields become tachyonic and indicate a decay of
the brane configuration. 

One way to get rid of the tadpoles associated with the geometrical, complex
structure moduli $U^i$ is to perform an orbifold twist on the
toroidal background (see next paragraph) such that some or
all of the complex structure moduli get frozen
\cite{Blumenhagen:2001te}. E.g.
in the ${\mathbb Z}_3$ 
orbifold all $U^i$ moduli are absent. However one is always
left with the dilaton tadpole. Since the potential has runaway
behavior in the dilaton direction, one is generically driven
to zero coupling. Therefore it is still an open problem to
cancel the dilaton tadpole for finite values of the string coupling constant
in non-supersymmetric models (or also after supersymmetry breaking in
supersymmetric models).

Next let us consider the action of the orientifold and of the orbifold
group,
where the spatial orbifold group is defined
by elements from ${\mathbb Z}_N$ (or ${\mathbb Z}_N\times {\mathbb Z}_M$). 
The latter 
are represented by the  $\theta$ (and $\omega$), 
describing discrete rotations on the 
compact coordinates $x^i,y^i$. This 
action restricts the compactification lattice
and fixes some of the internal parameter (\ref{moduli}) to discrete values.
The orientifold $O6$--planes describe the 
set of points which are invariant under the group 
actions $\Omega\overline\sigma$,
$\Omega\overline\sigma\theta^k$, $\Omega\overline\sigma\omega^l$ and
$\Omega\overline\sigma\theta^k\omega^l$. 
These planes are generated
by rotations of the real $x^j$ axes by $\theta^{-k/2}\omega^{-l/2}$.

The condition for tadpole cancellations in IIA
orientifold backgrounds in four space--time 
dimensions requires a system of $D6$ branes which has to respect the orbifold 
and orientifold 
projections.
In particular,  for consistency with 
the orbifold/orientifold group their orbifold/orientifold 
mirrors have to be introduced.
Hence any stack $a$ is 
organized in orbits, which 
represent an equivalence class $[a]$. For $N,M\neq 2$ the length 
of each orbit $[a]$ is at most $2NM$, but may be smaller, if e.g. 
stack $a$ is located along an 
orientifold plane. 
Stacks within a conjugacy class $[a]$ have 
non--trivial intersections among each other and 
w.r.t.  stacks from a different class $[b]$ 
belonging to the gauge group $G_b$. Without going further into
any details, it is appealing that
one can check in several examples \cite{Blumenhagen:2002wn} 
that the open string spectrum constructed following
these rules in the
toroidal ambient space
precisely agrees with the geometrical spectrum discussed
before, when considering the orbifold space as a limiting geometry
of a Calabi-Yau manifold.
The requirement of $R$--tadpole cancellation 
leads to some constraints on the number and location
of the $D6$-branes. 
To derive the tadpole conditions one has to compute the
Klein-bottle, annulus
and M\"obius strip amplitudes.
In the annulus amplitude  all open string
sectors contribute including those from open strings stretched between 
two branes belonging to the same equivalence class. It turns out to 
be convenient to define the following two quantities for any equivalence class
$[(n^i_a,m^i_a)]$ of D6$_a$-branes 
(here for the ${\mathbb Z}_3$-orbifold    \cite{Blumenhagen:2001te})
\begin{eqnarray}
       Z_{[a]}&=&{2 \over 3} \sum_{({n}^i_b,{m}^i_b)\in [a]}
                           \prod_{i=1}^3  \left( {n}^i_b +{1\over 2}\,
                         {m}^i_b \right) , \nonumber \\
       Y_{[a]}&=& -{1\over 2} \sum_{({n}^i_b,{m}^i_b)\in [a]}
                          (-1)^M\, \prod_{i=1}^3  
                         {m}^i_b  
\end{eqnarray}
where $M$ is defined to be odd for a mirror brane and otherwise even. The sums 
are taken over all the individual D6$_b$-branes that are elements of the
orbit $[a]$.       
If we introduce $K$ stacks of equivalences classes $[a]$  of branes,
then the RR-tadpole cancellation condition reads
\begin{equation}  \sum_{a=1}^K N_a\, Z_{[a]} =2 .
\end{equation}
Note, that the sum is over equivalence classes of D6-branes.
This equation is equivalent to the geometrical tadpole conditions
eq.(\ref{tadpoles}).

\vskip0.2cm
Non-supersymmetric toroidal
and orbifold models with SM-spectrum can be constructed in
various ways. E.g. in 
\cite{Blumenhagen:2001te}
 SM intersecting brane worlds 
on a 
${\mathbb Z}_3$-orbifold were explicitly constructed.
Also, specific examples of orbifold
intersecting brane world models with ${\cal N}=1$ supersymmetry in $D=4$
have been introduced in 
\cite{Cvetic:2001tj,Cvetic:2001nr,
Blumenhagen:2002gw,Honecker:2003vq,Honecker:2003vw}.
However to derive ${\cal N}=1$ MSSM models seems to be much more involved.

\vskip0.4cm\noindent
As a specific example of an intersecting
brane model on a Calabi-Yau
manifold let us display the construction
of the non-supersymmetric Standard Model on 
the Fermat quintic Calabi-Yau  
\cite{Blumenhagen:2002wn,Blumenhagen:2002vp}, which is defined
by the following equation in 
$\mathbb{CP}^4$:
\beqn
P(z_i) = z_1^5 + z_2^5 + z_3^5 + z_4^5 + z_5^5 = 0 \subset \mathbb{CP}^4\, .
\eeqn
The orientifold six-plane $\pi_{O6}$ is determined by the real counter part
of this equation:
\beqn
P(x_i) = x_1^5 + x_2^5 + x_3^5 + x_4^5 + x_5^5 = 0
\subset \mathbb{RP}^4\, .
\eeqn

Next we have to introduce the D6-branes being wrapped around the
supersymmetric 3-cycles. These are obtained by applying the
 $\mathbb{Z}_5^4$ symmetry, 
 $z_i\mapsto \omega^{k_i}z_i$,
$\omega=e^{2\pi i/5},\ k_i\in\mathbb{Z}_5$, on $\pi_{O6}$.
Specifically each  sLag 3-cycle can be characterized by four
integers, 
$\pi_a=\pi_{k_2,k_3,k_4,k_5}$, and is defined by the
following equation:
\beqn
x_1^5 + \Re(\omega^{k_2}z_2)^5 + \Re(\omega^{k_3}z_3)^5+\Re(\omega^{k_4}z_4)^5+
\Re(\omega^{k_5}z_5)^5
 = 0
\eeqn
This leads to a set of $5^4=625$ sLag $\mathbb{RP}^3$'s, calibrated with
$\Re(\prod_i \omega^{k_i} \Omega_3)$. Of these, 
125 sLag's are calibrated with respect to
$\Re(\Omega_3)$, i.e. they are supersymmetric with respect to
the orientifold plane $\pi_{O6}$. Therefore, for  supersymmetric
D-brane configurations only these 3-cycles must be used. On the other hand,
the 3-cycles, which are calibrated in the same way, 
have zero intersection with each other. Hence in order to get chiral
fermions only non-supersymmetric models can be built on
the quintic Calabi-Yau in this way.

The general intersection numbers are already computed in the paper by
\cite{Brunner:1999jq}. To engineer the standard model we
introduce { four stacks of D6-branes} with 
N$_a=3$, N$_b=2$ and N$_c=$N$_d=1$, corresponding to
the gauge group $G=U(3)\times U(2)\times U(1)^2$.
Finally we choose the following ``wrapping numbers" for the D6-branes on
the slag 3-ycles:
\beqn
&& \pi_a = \pi_c-\pi_d
-\vert 0,2,1,4 \rangle-\vert 0,3,4,1\rangle
, \nonumber\\ &&
\pi_b = \vert 0,3,1,1
\rangle , \nonumber\\ && \pi_c = \vert 1,4,3,4 \rangle +\vert 4,4,3,2
\rangle , \nonumber \\ && \pi_d = \vert 0,3,0,3 \rangle - \vert 2,0,3,4
\rangle
\eeqn
This produces the intersection numbers of the Standard Model with
three generations of quarks and leptons.
The anomaly-free, massless hypercharge is
\beqn
U(1)_Y = \frac{1}{3} U(1)_a - U(1)_c + U(1)\, .
\eeqn
Finally, 
an hidden sector is needed for tadpole cancellation.

\section{Some phenomenological issues of intersecting brane worlds}

\subsection{Gauge coupling unification}

One of the biggest successes of the MSSM is the apparent unification
of the
three gauge coupling constants.
Remember that at the weak energy scale
the three {Standard Model gauge couplings $g_s$, $g_w$ and $g_y$
have quite different values.
Extrapolating these couplings via
the {one-loop renormalization group equations}
\begin{eqnarray}
 {4\pi\over g_a^2(\mu)}=k_a{4\pi\over g^2_X
}+{b_a\over 2\pi} \log\left({\mu\over M_X}\right)+\Delta_a 
\end{eqnarray}
to {higher energies}, one finds that they all meet at
\begin{eqnarray}
M_X\simeq  2\cdot 10^{16}~{\rm GeV}, \quad  
\alpha_s=\alpha_w={3\over 5}\alpha_Y=\alpha_X\simeq{1\over 24},\label{gut}
 \end{eqnarray}
if the
light spectrum contains {just the MSSM particles}. 
This is in accord with for instance an {$SU(5)$ Grand Unified gauge 
group} at the GUT scale.

In string theory one has a {new scale, the string scale $M_s$}, 
so that it is natural
to relate $M_X$ to $M_s$. In the heterotic string one finds
\begin{eqnarray}
k_a={\rm level\ of\ SU(N_a)\ Kac-Moody\ algebra},
\end{eqnarray}
and the heterotic relation between the string and the Planck scale
was found to be \cite{Kaplunovsky:1987rp,Derendinger:1991hq}
\begin{eqnarray}
M_s\simeq g_{st}\cdot 0.058\cdot  M_{pl},
\end{eqnarray}
which, using $g_{st}\simeq 0.7$, leads to
$M_s\simeq 5\cdot 10^{17}~{\rm GeV}$.
Assuming a MSSM like low-energy string spectrum,
this discrepancy between $M_X$ and $M_s$  needs to be explained
by moduli-dependent {string threshold corrections} $\Delta_a$
\cite{Dixon:1990pc,Ibanez:1991zv,Ibanez:1992hc,Nilles:1995kb}
(or alternatively by heterotic M-theory
\cite{Witten:1996mz}).

Let us now study gauge coupling unification 
in  type II
or in type I D-brane mo\-dels 
 with gauge group $SU(3)\times SU(2)\times U(1)_Y$
\cite{Antoniadis:2000en,Blumenhagen:2003jy}
(see also \cite{Krause:2000gp}); here
each gauge factor comes with its {own gauge coupling},
which at string tree-level can be deduced from the
Dirac-Born-Infeld action 
\begin{eqnarray}
 {4\pi \over g_a^2}={M_s^3\, V_{a} \over (2\pi)^3\, g_{st}\, 
\kappa_a}\, ,\quad V_a=(2\pi)^3R_a^3\, .
\end{eqnarray}
with $\kappa_a=1$ for $U(N_a)$ and $\kappa_a=2$ for $SP(2N_a)/SO(2N_a)$.

By dimensionally reducing the type IIA gravitational action one can  
similarly express the {Planck mass} in terms of the stringy parameters
($M_{pl}=(G_N)^{-{1\over 2}}$){
\begin{eqnarray}
{M_{pl}^2}={8\, M_s^8\, V_{6} \over (2\pi)^6\, g^2_{st}}\, ,\quad
V_6=(2\pi)^6R^6\, .
\end{eqnarray}}
Eliminating the unknown string coupling $g_{st}$ gives{
\begin{eqnarray}
{1\over \alpha_a}={M_{pl} \over 2\sqrt{2}\, \kappa_a\, M_s} 
                      {V_a\over \sqrt{V_6}}.
 \nonumber
\end{eqnarray}}
Due to 
\begin{eqnarray}
{V_a\over \sqrt{V_6}}= \int_{\pi_a}  \Re(e^{i\phi_a}\widehat\Omega_3)
 \nonumber
\end{eqnarray}
the gauge coupling only depends on the {complex structure moduli}.

Since already at string tree-level the different type II,I gauge
couplings depend on different moduli fields and hence in general
may take arbitrary values, it seems that gauge coupling
unification is unnatural or only occurs merely as an accident.
However one can show that under some natural assumptions there will be
one relation between the three MSSM gauge couplings that 
is compatible with gauge coupling unification.
Namely consider 
gauge couplings of intersecting D-branes  in a model independent bottom
up approach where we ask for
the following three 
phenomenological requirements:
\begin{itemize}
\item
The SM branes mutually preserve ${\cal N}=1$ supersymmetry.

\item The intersection numbers realize a 3 generation MSSM.

\item The $U(1)_Y$ gauge boson is massless.

\end{itemize}
(Note that an orbifold model with only the MSSM states as massless string modes
is not yet constructed; however there  seems to be no fundamental obstacle
to get such a model using more general backgrounds.)
Without going into the
details \cite{Blumenhagen:2003jy}
one can show that these restrictions provide one relation between the internal
volumes $V_a$, i.e. between
the 3 gauge coupling constants:
\begin{eqnarray}
{1\over \alpha_Y}={2\over 3}{1\over \alpha_{s}} +
                                  {1\over \alpha_{w}}.\label{rela}
 \end{eqnarray}
This relation will allow for natural gauge coupling unification!
In fact, if one furthermore assumes that $\alpha_{s}=\alpha_{w}$,
this relation is identical to the $SU(5)$ GUT relation in eq.(\ref{gut}).

Let us further explore what are the allowed values of the string scale
which follow from gauge coupling unification using the relation
(\ref{rela}).
In the { absence
of threshold corrections} (see the discussion at the end of this section),
the one-loop running of the three gauge couplings is described
by the well known formulas
\begin{eqnarray}
{1\over \alpha_s(\mu)}&=&{1\over \alpha_s} + {b_3\over 2\pi} 
                   \ln\left({\mu\over M_s}\right) \nonumber \\
                   {\sin^2\theta_w(\mu)\over \alpha(\mu)}&=&{1\over \alpha_w} + 
                        {b_2\over 2\pi} 
                   \ln\left({\mu\over M_s}\right) \nonumber\\
                   {\cos^2\theta_w(\mu)\over \alpha(\mu)}&=&{1\over \alpha_Y} + 
                        {b_1\over 2\pi} 
                   \ln\left({\mu\over M_s}\right) ,
\end{eqnarray}
where $(b_3,b_2,b_1)$ are 
the one-loop beta-function coefficients for $SU(3)_c$,
$SU(2)_L$ and $U(1)_Y$, and
$\alpha$ is the electromagnetic fine structure constant.
Using the {tree level relation} 
(\ref{rela})
at the string scale yields 
\begin{eqnarray}
   {2\over 3}{1\over \alpha_{s}(\mu)}+{2\sin^2\theta_w(\mu)-
                1 \over \alpha(\mu)}={B\over 2\pi}\, 
     \ln\left({\mu\over M_s}\right) 
\end{eqnarray}
with
\begin{eqnarray}
  B={2\over 3}\,b_3+ b_2-b_1.
\end{eqnarray}
Now employing  the {measured Standard Model parameters}
\begin{eqnarray}
M_Z&=&91.1876\ {\rm GeV},\quad \alpha_s(M_Z)=0.1172, \nonumber \\
                     \ \alpha(M_Z)&=&{1\over
127.934}, \quad \, \, \sin^2\theta_w(M_Z)=0.23113\, , \nonumber \\
\end{eqnarray}
the resulting value of the unification scale
{only depends on the combination $B$} of the beta-function coefficients. 

For the { MSSM} spectrum
one has  $(b_3,b_2,b_1)=(3,-1,-11)$, i.e $B=12$ and
the unification scale is the usual {GUT scale}
\begin{eqnarray}
M_s=M_X=2.04\cdot 10^{16}\, {\rm GeV}.
\end{eqnarray}

\noindent
For
the individual {gauge couplings at the string scale} we get
\ba
\alpha_s(M_s)=\alpha_w(M_s)={5\over 3}\alpha_Y(M_s)=0.041,
\ea
which are just the supersymmetric {GUT scale} values with the Weinberg angle
being $\sin^2\theta_w(M_s)=3/8$.

\noindent
Assuming $g_{st}=g_X$, one obtains
for the {overall radius $R$} and the  {internal radii $R_s,R_w$}
\ba
M_s R=5.32, \quad M_s R_s=2.6, \quad M_s R_w=3.3.
 \ea
Of course it still remains to be checked if the parameters can be chosen
in such a way in a given, concrete model.

\vskip0.2cm
One can also investigate the gauge coupling unification in other classes
of models with different low energy spectrum. For example,
many models contain
 besides the chiral matter also additional
vector-like matter.
These states are  also localized on the intersection
loci of the $D6$ branes and also come with {multiplicity $n_{ij}$} with
$i,j\in\{a,b,c,d\}$. 

\noindent
One finds the following contribution to $B$
\ba
  B&=&{12}  - 2\, n_{aa} - {4}\, n_{ab} + 2\, n_{a'c} 
         + 2\, n_{a'd} - 2\, n_{bb}  +
             2\, n_{c'c} \nonumber \\
  &&  +  2\, n_{c'd}  + 2\, n_{d'd}. 
\ea
$B$ does not depend on the number of {weak Higgs} multiplets $n_{bc}$.

\vskip0.2cm
\underbar{Example A:} 

\vskip0.2cm
\noindent
If we have a model with a {second weak Higgs}
field, i.e.  $n_{bc}=1$,   we still
get $B=12$ but with 
\ba
(b_3,b_2,b_1)=(3,-2,-12).
 \ea
The {gauge couplings} 
"unify" at the scale
\ba
M_s=2.02\cdot 10^{16}{\rm GeV}.
 \nonumber
\ea
However they are not all equal at that scale
\ba
  \alpha_s(M_s)=0.041, \quad \alpha_w(M_s)=0.052, \quad
                  \alpha_Y(M_s)=0.028. \nonumber
\ea
The evolution of the gauge coupling constants for this class of models
is shown in the following diagram:

\begin{center}
\epsfig{file=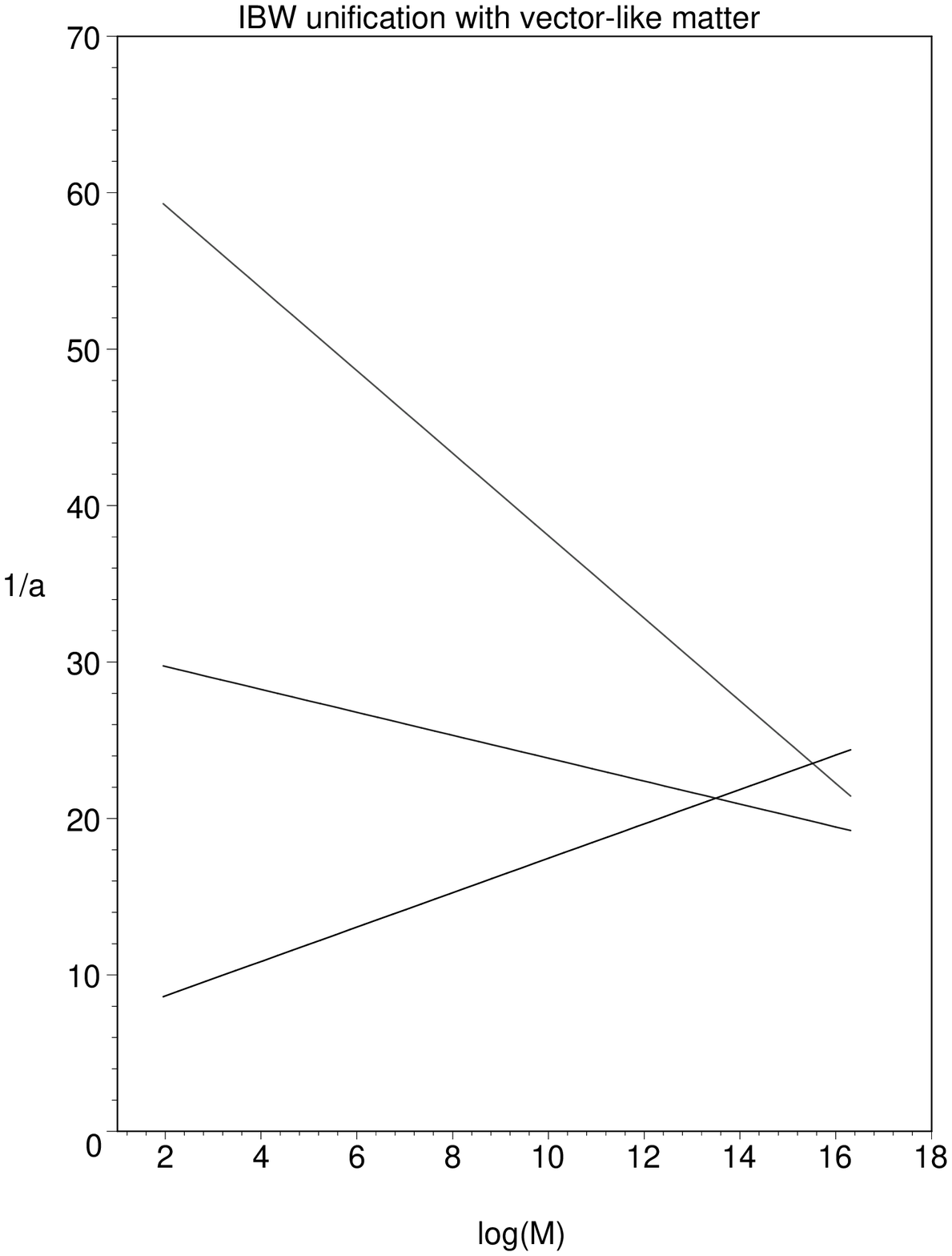,width=10cm}
\end{center}

\vskip0.2cm

\underbar{Example B: intermediate scale model} 

\vskip0.2cm
\noindent
For models with {gravity mediated supersymmetry breaking}
(hidden anti-branes) the string scale is naturally in the {intermediate
regime} $M_s\simeq 10^{11} {\rm GeV}$. 

\noindent
Choosing {vector-like} matter 
\ba
n_{a'a}=n_{a'd}=n_{d'd}=2, \quad  n_{bb}=1
\ea
leads to {$B=18$}.
The string scale turns out to be 
\ba
M_s=3.36\cdot 10^{11}{\rm GeV}.
\ea
The {running} of the couplings with 
\ba
(b_3,b_2,b_1)=(-1,-3,-65/3)
\nonumber
\ea
leads to the values  of the { gauge couplings} at 
the string scale
\ba
  \alpha_s(M_s)=0.199, \quad \alpha_w(M_s)=0.052, \quad
                  \alpha_Y(M_s)=0.045. \nonumber
\ea
Assuming $g_{st}\simeq 1$, one obtains for the radii
\ba
M_s R=230, \quad M_s R_s=1.7, \quad M_s R_w=3.3. 
\ea

\begin{center}
\epsfig{file=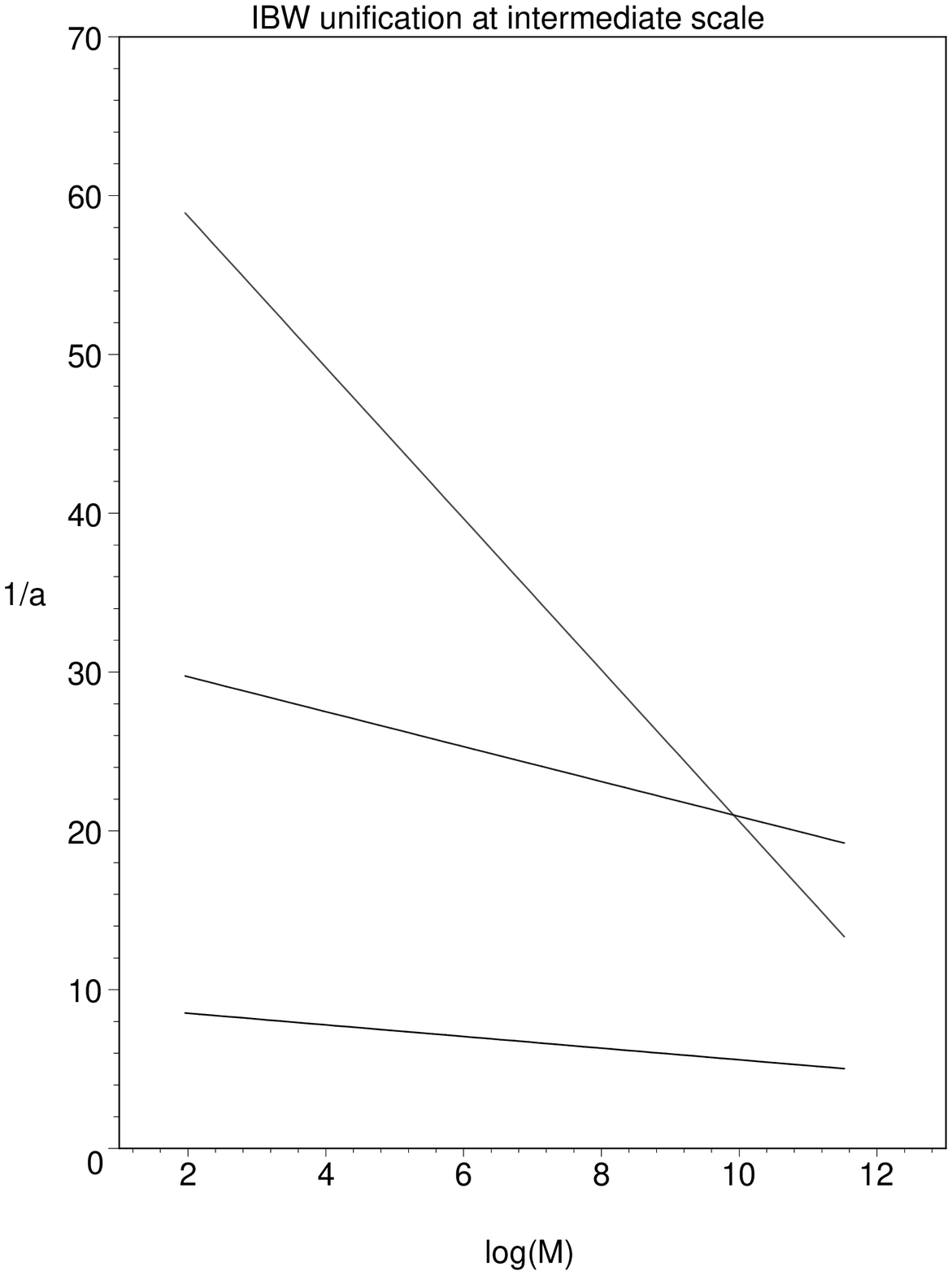,width=10cm}
\end{center}

\vskip0.2cm
\underbar{Example C: Planck scale model}

\vskip0.2cm
\noindent
Interestingly for {$B=10$} one gets
\ba
      {M_{s}\over M_{pl}}=1.24\sim \sqrt{\pi\over 2}. 
\ea
Choosing { vector-like} matter
\ba
n_{aa}=1,
\nonumber
\ea
the beta-function coefficients read 
\ba
(b_3,b_2,b_1)=(0,-1,-11).
\nonumber
\ea
The {couplings} at the string  scale turn out to be
\ba
  \alpha_s(M_s)=0.117, \quad \alpha_w(M_s)=0.043, \quad
                  \alpha_Y(M_s)=0.035 
\ea
leading to $\sin^2 \theta_w(M_s)=0.445$.
For the {scales}  of the overall Calabi-Yau volume and the 3-cycles
we obtain
\ba
M_s R=0.6, \quad M_s R_s=1.9,\quad M_s R_w=3.3 . 
\ea

\begin{center}
\epsfig{file=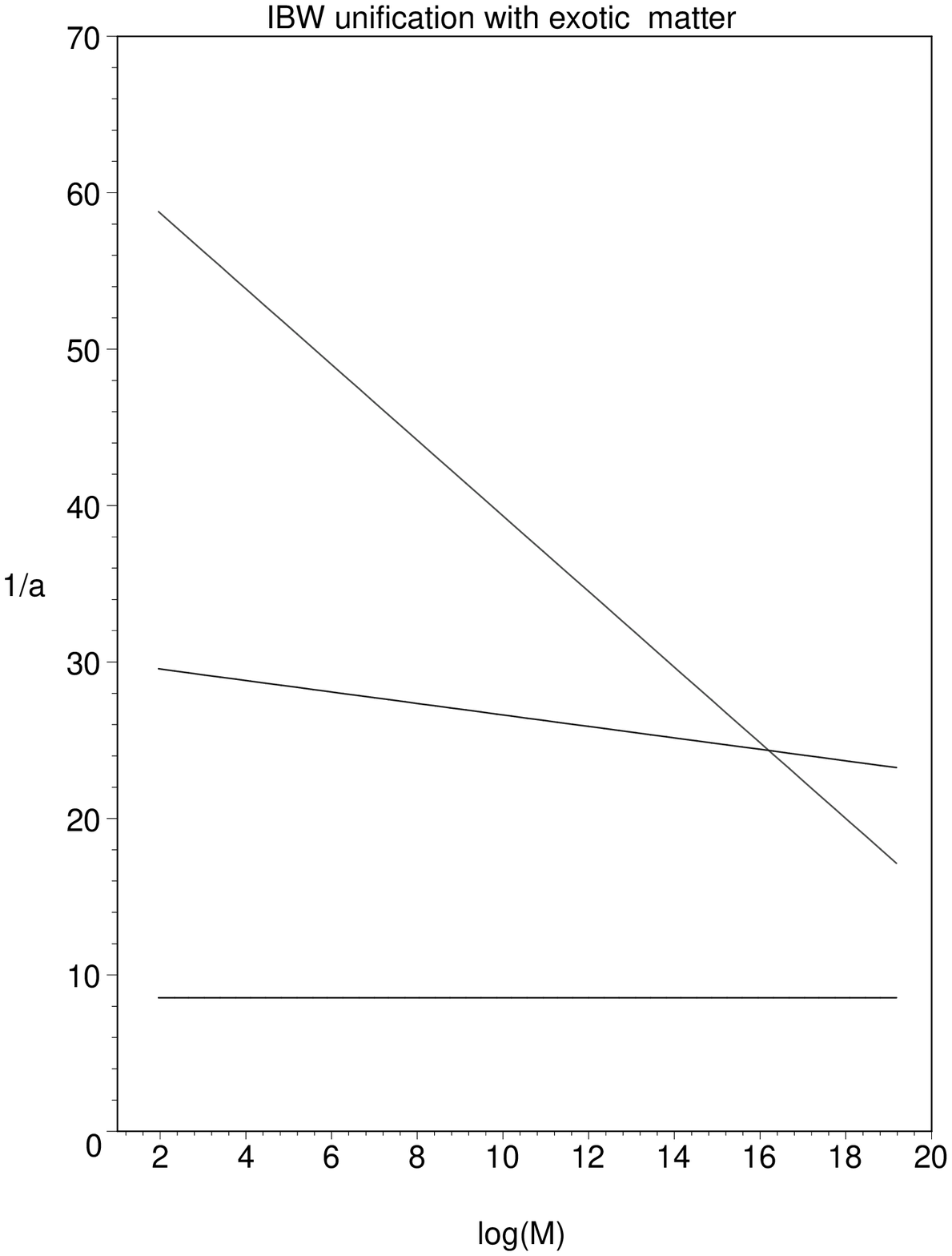,width=10cm}
\end{center}

\vskip0.3cm\noindent
So far we have neglected the stringy, one-loop gauge threshold
corrections $\Delta_a$ (to the gauge group $G_a)$
in the renormalization group equations
of the gauge coupling constant. We do not give a phenomenological
investigation of these threshold corrections, but sketch how they
are computed in intersecting brane world models.
Note however that the one--loop gauge 
threshold corrections $\Delta_a$, which take 
into account Kaluza--Klein and winding states from the internal dimensions and 
the heavy string modes, may change the
unification picture discussed before. 
For certain regions in moduli 
space these corrections may become huge and 
thus have a substantial impact on the 
unification scale.

In the type IIA picture with intersecting D6-branes these threshold
correction $\Delta_a$ will depend on the homology classes on the 3-cycles 
(open string parameters) and also
on the closed string geometrical moduli.
In toroidal models these corrections will be given in terms of the
wrapping numbers $n_a^j,m_a^j$ and the radii $R_i^j$ of the torus. 
All D6-branes have in common their four--dimensional (non--compact) 
world volume. Hence their gauge fields are located on parallel 
four--dimensional subspaces, which may be separated 
(in the cases $I_{ab}^j\neq 0$ and $I_{aa'}^j\neq 0$) in the 
transverse internal 
dimensions. One--loop corrections to the gauge couplings are realized through
exchanges of open strings in that
transverse space. The open string charges $q_a,q_b$ at their ends couple 
to the external gauge fields sitting on the branes.
Only annulus and M\"obius diagrams contribute, as torus and 
Klein bottle diagrams refer to closed string states.
In \cite{Lust:2003ky} the one--loop corrections to the gauge couplings 
were computed by the background field
method: one turns on 
a (space--time) magnetic field, e.g. $F_{23}=B Q_a$ in the $X^1$--direction 
and determines the dependence of the open string partition function on that
field.
Here, $Q_a$ is an appropriately normalized generator of the gauge group $G_a$ 
under consideration.
This leads to the so-called gauged open string partition function.
The second order of an expansion w.r.t.  $B$ of the gauged partition function  gives 
the relevant piece for the one--loop gauge couplings.

In the following we will omit all details of the calculations,
referring  the reader to the paper \cite{Lust:2003ky};
instead we give only the end results for the one loop threshold corrections:

\vskip0.2cm
\noindent (i) ${\cal N}=4$ sectors: $\Delta_a=0$.

\vskip0.2cm
\noindent (ii) ${\cal N}=2$ sectors:

\vskip0.2cm
\noindent
If the stacks $a$ and $b$ preserve ${\cal N}=2$ supersymmetry, i.e. they 
are parallel within some torus  $T_2^i$, we obtain for the gauge group $G_a$:
\begin{equation}{
\Delta^{\rm N=2}_{ab}\sim b_{ab}^{N=2}\ \ln (T^i_2V_a^i|\eta(T^i)|^4)+{\rm
  const.}\ ,}
\end{equation}
with the wrapped brane volume 
\begin{equation}
V_a^i=\frac{1}{U_2^i}|n_a^i+U^im^i_a|^2\, ,
\end{equation}
and the K\"ahler modulus $T^i$ defined in eq.(\ref{moduli}).

\vskip0.2cm
\noindent (iii) ${\cal N}=1$ sectors:

\vskip0.2cm
\noindent
In the case that the branes from $a$ and $b$ 
preserve ${\cal N}=1$ supersymmetry,
the one--loop correction  to the gauge coupling of $G_a$ takes the form:
\begin{equation}{
\Delta^{\rm N=1}_{ab}=-
b_{ab}^{N=1}\ \ln \frac{\Gamma(1-\frac{1}{\pi}\Phi_{ba}^1)\ 
\Gamma(1-\frac{1}{\pi}\Phi_{ba}^2)\ \Gamma(1+\frac{1}{\pi}\Phi_{ba}^1+
\frac{1}{\pi}\Phi_{ba}^2)}{\Gamma(1+\frac{1}{\pi}\Phi_{ba}^1)\ 
\Gamma(1+\frac{1}{\pi}\Phi_{ba}^2)\
\Gamma(1-\frac{1}{\pi}\Phi_{ba}^1-\frac{1}{\pi}\Phi_{ba}^2)}  \ .}
\end{equation}
This expression depends on the closed string moduli of the underlying
toroidal geometry, since the
the difference of the angles $\Phi^j_a$ and $\Phi^j_{a'}$ are 
related to the radii through:
\begin{equation}{
\coth(\pi v_{aa'}^j)=i\cot(\Phi_{a'}^j-\Phi_a^j)=
i\ \frac{n_a^jn_{a'}^j\frac{R_1^j}{R_2^j}+m_a^jm_{a'}^j\frac{R_2^j}{R_1^j}}
{n^j_a m_{a'}^j-
n^j_{a'} m^j_a}\ .}
\end{equation}
Note that this type of moduli dependence of the ${\cal N}=1$ threshold
functions in
intersecting brane world models is completely new, as in heterotic string
compactifications the ${\cal N}=1$ thresholds are moduli
independent constants \cite{Dixon:1990pc}.

In addition one should emphasize that in supersymmetric brane
world models
there are no UV divergences in the one-loop ${\cal N}=2,1$
thresholds. In fact one can show that cancellation of the vacuum
RR tadpoles implies that in supersymmetric
models also all RR and NS tadpoles are absent
in the one-loop 2-points functions for the gauge
couplings, i.e. in the gauged open string partition
functions. This proves the finiteness of the supersymmetric one-loop
gauge thresholds in the considered class of orbifold models.

\subsection{Proton decay}
At the end of the paper let us also discuss briefly
some aspects of proton decay
in intersecting brane world models. 
Here we are following the work of \cite{Klebanov:2003my}.  Recall that in 
supersymmetric GUT field theories
proton decay can occur due to { dimension 5 operators $\int d^2\theta Q^3L$,
i.e. exchange of SUSY particles $S$,}
or can occur due to { dimension 6 operators $\int d^4\theta Q^2\tilde Q^*
\tilde L^*$,
i.e. exchange of heavy gauge vector bosons $X$.
In string theory (or in M-theory) has  to 
deal 
in addition with the exchange of  {an infinite tower of KK states,}
as shown in the following figure:

\vskip0.8cm

\begin{center}
\epsfig{file=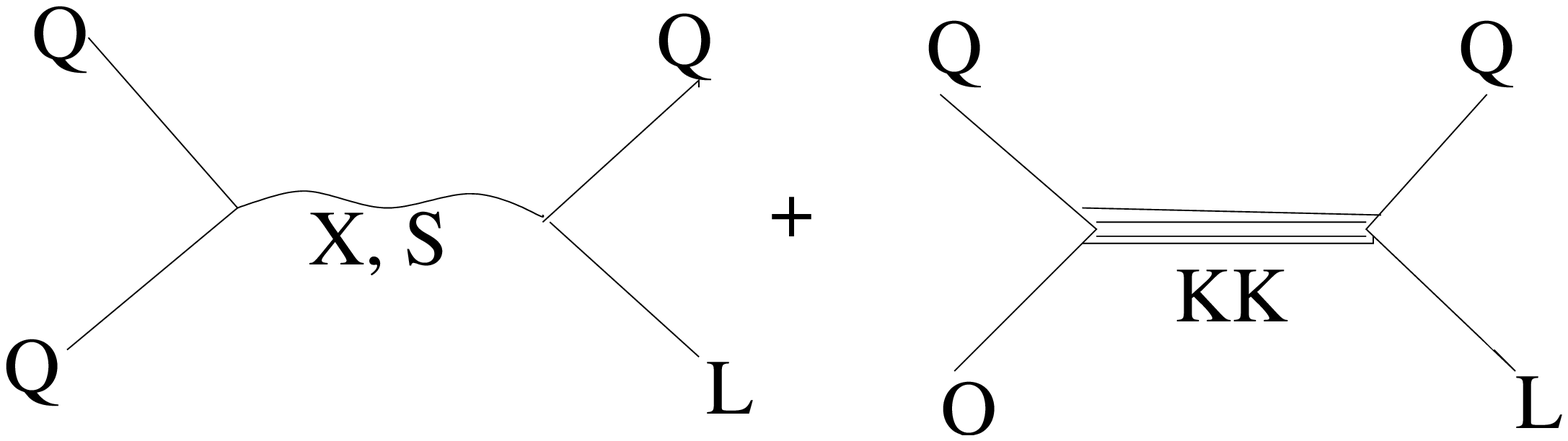,width=10cm} 
\end{center}

{Due to the KK 
exchange one expects an 
enhancement of the proton decay amplitude by an universal factor 
$\alpha_{GUT}^{-1/3}$.}
However to know the precise numerical coefficient one has to make
a model dependent calculation.

First consider shortly 
intersecting D6-brane models with SM gauge
group{ $G=SU(3)\times SU(2)\times
  U(1)_Y$. Here
baryon number is an anomalous $U(1)$ gauge symmetry, and hence there
is a 
global $U(1)_B$ symmetry. Therefore proton decay is completely forbidden
in this class of models.

Next turn to intersecting D6-brane  models with{ $SU(5)$-GUT gauge group.
This can be realized by a stack of 
five D6-branes plus their orientifold
mirrors D6'.
The massless spectrum is given by
$SU(5)$ gauge bosons and
{\bf $10+{\overline 10}$} matter from open strings at the intersection
of D6,D6'-branes.
Note that 
there are no fundamental matter fields
in the {\bf $5$}-representation in this brane set up.
Now one 
can compute the { open string disc amplitude} {\bf $10^2{\overline 10}^2$}
($p\rightarrow \pi^0e^+_L$):
\begin{eqnarray}
A_{st}&=&\pi\alpha'g_sI(\Phi_1,\Phi_2,\Phi_3)=
{\alpha_{GUT}^{2/3}
L^{2/3}g_s^{1/3}I(\Phi_1,\Phi_2,\Phi_3)\over 4\pi M_{GUT}^2}\, ,
\nonumber\\
I&=&\int_0^1{dx\over x(x-1)}\prod_{i=1}^3\sqrt{\sin(\pi\Phi_i)}
\lbrack F(\Phi_i,1-\Phi_i;1;x) \nonumber\\
&{~}&F(\Phi_i,1-\Phi_i;1;1-x)\rbrack^{-1/2}\simeq 7-11\, .
\end{eqnarray}
After having performed this string calculation we have to compare the
field theory proton lifetime  with the string theory lifetime:

\noindent
{Field theory:}
\begin{eqnarray}
\tau_p=1.6\times 10^{36}{\rm y}\Biggl({0.4\over \alpha_{GUT}}\Biggr)^2
\Biggl({M_X\over 2\times 10^{16}{\rm GeV}}\Biggr)^4\simeq 
1.6\times 10^{36}{\rm y}\, .
\end{eqnarray}

\noindent
{String theory:}
\begin{eqnarray}
\tau_{p,s}=
{(0.037 ~L^{2/3}Ig_s^{1/3})^{-2}}\tau_p
\, .
\end{eqnarray}

\vskip0.2cm
\noindent
This comparison displays the following two new enhancement factors:

\noindent
(i) The contribution of the KK-states leading to
a 1-loop threshold factor $L\simeq 8$
(being obtained from M-theory on a $G_2$-manifold \cite{Friedmann:2002ty}).

\noindent
(ii) The tree level string disc amplitude: $I\simeq 7-11$.

\noindent
Combining these two factors with $g_s={\cal O}(1)$
one learns that
no substantial enhancement of the proton lifetime is going to appear.
Note also that in order to
get some further informations on the
proton life time
in D-brane models also dimension 5 operators should be taken into account.

\section{Conclusion}

In this paper we have reviewed several aspects of intersecting brane world 
models. The main aspect was on the constructive part
of these models. As we have seen, 
it is possible to derive the SM spectrum from this kind of D-brane
constructions, at least in non-supersymmetric brane embeddings
into the compact space.
We also discussed how gauge coupling unification can be achieved, including
some discussion on the computation of one-loop gauge threshold
corrections; finally the computation of
the proton decay amplitude, which is based
on the calculation of an open string four-point amplitude
at the disc level, was briefly presented.

\vskip0.2cm
However there are several open problems in type I intersecting
brane world models which should be addressed in the future:

\begin{itemize}

\item
One of the main challenge remains to construct realistic supersymmetric
intersecting brane
world models with the chiral spectrum of the MSSM.

\item An important problem is to get more informations
on the low-energy effective action of intersecting brane world models.
One way to obtain the effective action is to compute open/closed
string scattering amplitudes and extract from them the relevant low-energy
couplings. In this way the moduli dependent gauge couplings, and the
matter field K\"ahler potential can be derived \cite{paper}.
Furthermore these results can be used to compute the soft supersymmetry
breaking parameters in the low-energy effective action after
assuming some mechanism for supersymmetry breaking 
(see also \cite{Kors:2003wf}).

\item Another interesting field of research is to combine D-brane
constructions with flux compactifications (see the talk by G. Dall'Agata
\cite{gd}).
This has the advantage that due to the fluxes some more moduli
could be fixed dynamically, which are so far left untouched
by the D-branes. In addition non-vanishing fluxes may provide a way for
supersymmetry breaking.
E.g. in \cite{Blumenhagen:2003vr,Cascales:2003zp}
a scenario of D9-branes with open string F-flux
(the T-dual mirror configuration to D6-branes at angles) plus
some Ramond and NS 3-form flux was considered (see also
\cite{Behrndt:2003ih} for a type IIA description).
Here the D9-brane potential fixes some of the K\"ahler moduli
fields, where the 3-form fluxes stabilize some of the complex structure
moduli.

\item
So far we have completely neglected the back-reaction of the D6-branes
(or of non-vanishing background fluxes) on the space-time geometry.
In general due to this back-reaction the internal space will not be
any longer a Ricci-flat Calabi-Yau manifold. In the case of
${\cal N}=1$ supersymmetry it will rather be a six-manifold
with a connection with torsion, namely a space with a specific
kind of $SU(3)$ group structure (see the talk by G. Dall'Agata \cite{gd}).
One way to get some informations on these new spaces is to consider
the M-theory lift of supersymmetric D-brane models. 
For intersecting D6-branes plus O6-planes the M-theory background
will become purely geometrical, namely a certain singular 7-manifold
with $G_2$ holonomy  \cite{Atiyah:2001qf,Cvetic:2001kk,Behrndt:2002xm}. 
Thus our ${\cal N}=1$ gauge threshold function is possibly
related to the recently calculated Ray--Singer torsion of 
singular $G_2$ manifolds \cite{Friedmann:2002ty}.
Hence on the M-theory side the threshold function
is expected to be mapped to a topological quantity, 
most likely to the elliptic genus of 
the singular $G_2$ compactification manifold.

\item
Of course, the question of how to select the correct ground state
in string theory still remains an open problem
(for some thoughts on the statistics of string vacua see
\cite{Douglas:2003um}).

\end{itemize}
 
\section*{Acknowledgments}
I would like to thank the organizers of the Copenhagen workshop to provide
a very stimulating and pleasant athmosphere and for their hospitality.
In addition
I would like to thank my collaborators Ralph Blumenhagen, Volker Braun, Lars
G\"orlich, Boris K\"ors, Tassilo Ott and Stephan Stieberger
for the very pleasant collaboration on the material presented in this paper.
This work is supported in part by the European Community's Human Potential
Programme under the contract HPRN-CT-2000-00131 Quantum Spacetime.

\end{document}